\documentclass{jfm}

\usepackage{graphicx}
\usepackage{cancel}
\usepackage{newtxtext}
\usepackage{newtxmath}
\usepackage{natbib}
\usepackage{hyperref}
\hypersetup{
    colorlinks = true,
    urlcolor   = blue,
    citecolor  = black,
}

\newcommand{\RomanNumeralCaps}[1]
\linenumbers

\title{Instability of a dusty shear flow}

\author{Anu V. S. Nath\aff{1},
  Anubhab Roy \aff{1}
 \and M. Houssem Kasbaoui \aff{2} \corresp{\email{houssem.kasbaoui@asu.edu}}}

\affiliation{\aff{1}Department of Applied Mechanics and Biomedical Engineering, Indian Institute of Technology Madras, Chennai 600036, India
\aff{2}School for Engineering of Matter, Transport and Energy, Arizona State University,
Tempe, AZ 85281, USA}

\begin{document}
\maketitle

\begin{abstract}
We study the instability of a dusty simple shear flow where the dust particles are distributed non-uniformly. A simple shear flow is modally stable to infinitesimal perturbations. Also, a band of particles remains unaffected in the absence of any background flow. However, we demonstrate that the combined scenario -- comprising a simple shear flow with a localised band of particles -- can exhibit destabilisation due to their two-way interaction. The instability originates solely from the momentum feedback from the particle phase to the fluid phase. Eulerian-Lagrangian simulations are employed to illustrate the existence of this instability. Furthermore, the results are compared with a linear stability analysis of the system using an Eulerian-Eulerian model. Our findings indicate that the instability has an inviscid origin and is characterised by a critical wavelength below which it is not persistent. 
We have observed that increasing particle inertia dampens the unstable modes, whereas the strength of the instability increases with the strength of the coupling between the fluid and particle phases.
\end{abstract}

\begin{keywords}
Authors should not enter keywords on the manuscript, as these must be chosen by the author during the online submission process and will then be added during the typesetting process (see \href{https://www.cambridge.org/core/journals/journal-of-fluid-mechanics/information/list-of-keywords}{Keyword PDF} for the full list).  Other classifications will be added at the same time.
\end{keywords}

{\bf MSC Codes }  {\it(Optional)} Please enter your MSC Codes here

\section{Introduction}
\label{sec1}
Particle-laden flows are prevalent in various environmental, astrophysical, and industrial settings. In the environmental context, these flows play pivotal roles in shaping the landscape around us through sediment transport \citep{burns2015sediment}, influencing weather patterns via clouds \citep{pruppacher1998microphysics,shaw2003particle}, sea-spray \citep{veron2015ocean}, volcanic eruptions \citep{bercovici2010two}, and gravity currents \citep{necker2005mixing}. Astrophysical scenarios include particle-laden flows in cosmic dust clouds, stellar winds, and the transport of interstellar dust particles in accretion disks and proto-planetary disks, which are crucial for forming and evolving planetary systems \citep{youdin2005streaming,fu2014effects,homann2016effect}. In industrial and engineering applications, particle-laden flows arise in processes like spray drying \citep{straatsma1999spray,birchal2006spray}, powder handling \citep{baxter2000simulation}, and fluidized bed reactors \citep{bi2000state}. Particle-laden flows typically involve multiple components, with at least a carrier phase and a dispersed phase, leading to physics occurring at multiple scales. In natural scenarios, the carrier flows are often turbulent, and the suspended particles are advected and sheared by this turbulence. Modelling particle-laden flows within a sufficiently dilute limit often involves considering the momentum exchange from the fluid to the particles while neglecting feedback from particles to the fluid (one-way coupling). However, this feedback is significant, especially when the mass fraction of particles to fluid is on the order of unity (e.g., dusty gas flows or water droplets in the air), as it can introduce new dynamics into the system. In this study, we investigate a novel instability in a particle-laden simple shear flow arising from two-way coupling, where the feedback force from particles to fluid is considered.

The non-uniform distribution of particles, also known as particle segregation or particle banding, is observed in particle-laden flows across diverse engineering and environmental contexts. It can occur due to various segregation mechanisms such as preferential clustering, differential settling velocities, turbulent dispersion, and particle-particle interactions. For example, in turbulent flows, inertial particles can cluster preferentially in regions of high strain and lower vorticity, forming regions with high and low particle concentration \citep{maxey1987gravitational,bec2005multifractal,fiabane2012clustering}. This phenomenon can arise solely from one-way coupling, and no feedback force is required. In pneumatic conveying systems transporting granular materials, such as powders or grains, bands of particle-rich and particle-deficient regions can form (known as rope formation) due to agglomeration and flow dynamics \citep{klinzing2011pneumatic,huber1994characterization,lain2013characterisation,zhang2023coarse}. In turbulent wall-bounded flows carrying suspended particles, such as industrial pipelines transporting slurries, particle segregation can occur due to differential settling velocities, turbulent dispersion and thermophoresis. This segregation can lead to the formation of particle-rich layers near the bottom or walls of the flow channel, with particle-deficient regions in the core of the flow \citep{marchioli2003direct,sardina2012wall,picano2009spatial}. The particles form extremely long clusters, called ropes, and align preferentially with the low-speed turbulent streaks, contributing to their stabilization and suppression of bursting \citep{dave2023mechanisms}. Despite the additional stresses resulting from particles, the alteration of near-wall coherent structures results in a notable decrease in Reynolds shear stresses and partial relaminarization of the near-wall flow. Fluidized bed reactors used in chemical engineering often exhibit particle banding phenomena \citep{harris1994solitons,liu2016cfd,gilbertson2001segregation}. Observations of sediment-laden river flows and estuarine environments have revealed the formation of bands of sediment deposition influenced by flow dynamics, sediment transport mechanisms, coagulation and channel morphology \citep{gibbs1986segregation,sondi1995sedimentation,ogami2015dynamic}. These sediment bands play a crucial role in shaping riverbeds, deltas, and coastal environments.

In astrophysical accretion disks, such as those around young stars or black holes, there are regions where gas and dust particles orbit around a central object. The dust-gas system exhibits Keplerian motion, alongside radial and azimuthal drifts between the dust and gas. When the gas and dust move at slightly different velocities, this disparity can result in a relative drift between the two components. This relative motion creates a shearing force that can amplify small perturbations/disturbances in the dust distribution - known as streaming instability \citep{youdin2005streaming,chiang2010forming}. The name ``streaming instability" reflects the differential streaming motion between gas and dust particles within the disk. The instability arises from the relative drift between the gas and dust phases, a universal consequence of radial pressure gradients. Growth occurs even though the two components interact only via dissipative drag forces. Streaming instability exhibits growth rates that are slower than dynamical time scales but faster than drift time scales. As a result, dust particles start clumping together, even without self-gravity, forming dense structures or bands. Thus, the dust particles can localise, leading to additional dynamics or instabilities due to the Keplerian shear (as we see in this paper) and self gravity. Thus, streaming instability is crucial in various astrophysical processes, including planetesimals' formation and dust grains' growth in protoplanetary disks.

The hydrodynamic stability characteristics of particle-laden flows can be altered by modifying existing instabilities or generating new types of instabilities, as one considers the feedback from particles. Early studies by \cite{kazakevich1958investigations,sproull1961viscosity} observed a notable reduction in the resistance coefficient when dust was added to the air flowing turbulently through a pipe. It was thought that adding particles altering the effective viscosity led to this modification; however, this contradicts Einstein's formula for the suspension viscosity. \cite{saffman1962stability} was among the first to provide an analytical model for studying the stability of a dusty planar flow. He proposed that inertial particles extract energy from turbulent fluctuations, thereby damping them. He modelled both particle and fluid phases as a continuum using a two-fluid model. The momentum exchange between both phases is accounted for using a linear Stokes drag. The modal analysis ultimately resulted in a modified Orr-Sommerfeld equation for uniformly distributed particles. \cite{saffman1962stability} deduced that finer particles with low inertia could induce destabilization due to a reduction in effective kinematic viscosity. Although the effect of particles on the viscosity of dusty gas is negligible, it effectively increases the gas density, thereby reducing the kinematic viscosity. Conversely, coarser particles with high inertia could lead to stabilization through dissipation by Stokes drag. He concluded that dust merely alters the waves present in a clean gas and may not introduce any additional instabilities. Subsequent studies have numerically solved the modified Orr-Sommerfeld equation for various base flows, confirming Saffman's conclusions \citep{michael1964stability,asmolov1998stability,tong1999two,klinkenberg2011modal,sozza2022instability}. \cite{sozza2020drag,sozza2022instability} investigated a dusty Kolmogorov flow and demonstrated that increasing the particle mass loading reduces the amplitude of the mean flow and turbulence intensity. They observed that turbulence suppression is more pronounced for particles with smaller inertia. The study concluded that while inertia significantly influences particle dynamics, its impact on flow properties is negligible compared to mass loading. 

Notably, Saffman's analysis does not account for gravitational effects. Including gravity can introduce buoyancy effects that may destabilize the flow more easily \citep{herbolzheimer1983stability,shaqfeh1986effects,borhan1988sedimentation}. For example, \cite{magnani2021inertial} investigated dusty Rayleigh-Taylor turbulence and found that the interface between the two phases becomes unstable in the presence of gravity forces, evolving into a turbulent mixing layer that broadens over time. However, in a particle-laden Rayleigh–Bénard  system \citep{prakhar2021linear}, it was observed that particles tend to inhibit the onset of natural convection, thereby stabilizing the system. This is because particles act as a distributed drag force and heat source in the fluid, similar to a porous medium. Saffman's analysis, also, focusing solely on uniform particle distribution, overlooks the potential effects of non-uniform distributions. A study by \cite{narayanan2002temporal} has demonstrated the presence of additional unstable modes under large Stokes numbers and high mass loading conditions in a particle-laden mixing layer where the particle distribution is localised. Additionally, investigations utilising non-uniform particle distributions have highlighted the emergence of novel instabilities \citep{senatore2015effect,warrier2023stability}.

The non-uniform distribution of inertial particles arises naturally in vortical flows due to their preferential accumulation. As noted earlier, it has long been recognized that inertial particles tend to be centrifuged from vortical regions and cluster in regions of high strain \citep{maxey1987gravitational}, a phenomenon known as preferential accumulation. A numerical investigation by \cite{shuai2022accelerated} demonstrated that in a Lamb-Oseen vortex, inertial particles are expelled from the vortex core, forming a ring-shaped cluster and a void fraction bubble that expands outward. However, without accounting for the two-way interaction, the vortex would decay slowly due to viscous dissipation. When the two-way coupling is considered, it is observed that the feedback from clustered particles flattens the vorticity distribution and leads to an accelerated vortex decay. It is noted that as the inertia of the particles increases, the vorticity decays even more rapidly. A follow-up numerical study by \cite{shuai2022instability} on a particle-laden Rankine vortex revealed that the system becomes unstable due to the two-way interaction, which is also validated by analytical linear stability analysis. The feedback force from the particles triggers a novel instability, which can cause the breakdown of an otherwise resilient vortical structure. In the context of the merging of a pair of co-rotating vortices laden with inertial particles, it has been shown that \citep{shuai2024merger} the feedback force from particles significantly alters the monotonic merging behaviour as observed without feedback. The vortices push apart for a while due to a net repulsive force from particles ejected from the vortex cores, thus delaying their merging.

Apart from modal instability, the interplay between particle inertia and shear from the base flow can lead to transient growth of perturbations in particle-laden flows via non-modal growth mechanisms such as the Orr mechanism or the `lift-up' effect. Performing a non-modal analysis, \cite{klinkenberg2011modal} showed that transient growth in a particle-laden channel flow increases proportionally with the particle mass fraction. Similar non-modal instabilities have been observed and studied in dusty gas flows, such as the Blasius boundary layer with a localized dust layer \citep{boronin2014modal}, plane channel suspension flow with a Gaussian layer of particles \citep{boronin2016nonmodal}, and stably stratified Blasius boundary layer flow \citep{parente2020modal}. Additionally, the inclusion of gravitational effects in a simple shear flow has been shown to alter the uniformity of particle distribution, leading to the formation of local particle clusters and, subsequently, to transient growth \citep{kasbaoui2015preferential}.

The previously mentioned studies mostly used an Eulerian-Eulerian model for particle-laden flows. However, there are three primary methods for modelling particle-laden flows: (i) Eulerian-Eulerian modelling, (ii) Eulerian-Lagrangian modelling, and (iii) fully resolved simulations. The Eulerian-Eulerian method \citep[see][]{jackson2000dynamics,drew2006theory} treats both the particle and fluid phases as interpenetrating continua in an Eulerian framework, assuming particles are small ($\Delta x \gg d_p$) and sufficiently densely distributed to be treated as a fluid continuum. While computationally more affordable, this method may introduce errors due to the difficulty of modelling terms in Eulerian-Eulerian methods that require closure. In contrast, the Eulerian-Lagrangian method treats the fluid phase as a continuum within an Eulerian framework while representing the particle phase as discrete entities tracked individually in a Lagrangian manner. Here, the fluid phase is solved at a coarser scale, typically with $\Delta x$ a few times larger than $d_p$, resulting in a scalable and cost-effective approach. However, empirical coupling between the particle and fluid phases introduces some approximation errors. Few studies that considered Eulerian-Lagrangian modelling and stability of particle-laden flows are \cite{meiburg2000vorticity,richter2013momentum,senatore2015effect,kasbaoui2015preferential,wang2019modulation,kasbaoui2019turbulence,pandey2019clustering,shuai2022instability}. The fully resolved method, for example, realised by the immersed boundary method \citep[see][]{uhlmann2005immersed,breugem2012second,kempe2012improved,dave2023volume,Kasbaoui2024high-fidelity}, is employed when the grid spacing $\Delta x$ is significantly smaller than the particle diameter $d_p$. While offering high accuracy, this method requires extensive resolution, making it costly and impractical for large-scale applications. Thus, in this study, we only employ the Eulerian-Lagrangian and Eulerian-Eulerian methods, the details of which are discussed respectively in \S \ref{sec2p1} and \S \ref{sec4p1}.
\begin{figure}
    \centering
    \includegraphics[width = 0.8\linewidth]{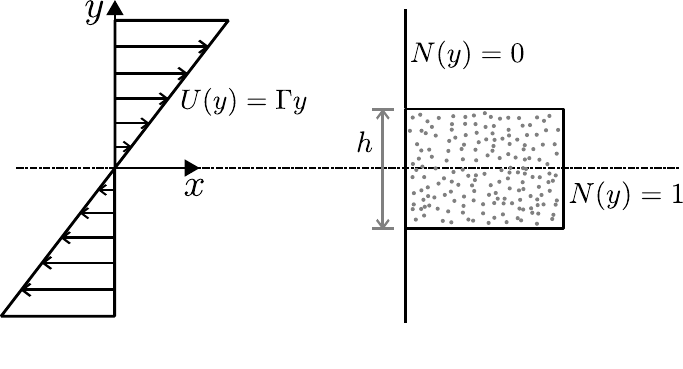}
    \caption{Schematic showing the configuration studied here: an unbounded simple shear flow (with a shear rate $\Gamma>0$) passing over a band of particles. The particles of uniform size are randomly distributed within a band of width $h$ with equal probability, forming a top-hat distribution.}
    \label{fig1}
\end{figure}

Here, we investigate the stability of a dusty simple shear flow with non-uniformly distributed particles. In the absence of particles, a simple shear flow is modally stable to infinitesimal perturbations. Similarly, a non-uniform particle distribution without any background flow remains unaffected. Thus, each system under consideration is linearly stable when inspected in isolation. However, when considering the combined system (see the schematic in figure \ref{fig1}), where a simple shear flow is superimposed on a band of particles, the background flow advects the particles, and the feedback from the particles induces an instability in the system. This paper reveals that this seemingly simple yet crucial particle-laden system exhibits a novel type of instability when incorporating two-way coupling. In \S \ref{sec2}, we demonstrate the presence of this novel instability through Eulerian-Lagrangian numerical simulations. The mechanism underlying the genesis of this new instability is described in \S \ref{sec3}. An analytical study of the system is carried out in \S \ref{sec4} using an Eulerian-Eulerian framework. Following the approach of \cite{saffman1962stability}, linear stability analysis is employed to establish the existence and modal nature of the instability. \S \ref{sec5} offers a comparative analysis between the Eulerian-Lagrangian and Eulerian-Eulerian results. Finally, we conclude in \S \ref{sec6}.


\section{Evidence of instability in a two-way coupled particle-laden shear flow}
\label{sec2}
In this section, we present a novel instability occurring in a particle-laden simple shear flow. Along with the momentum transport from fluid to particle phase, we consider the feedback from the particle to fluid phase (two-way coupling), which is crucial for the instability to occur. We employ an Eulerian-Lagrangian (EL) method to illustrate this instability. Below, we describe the method used:



\subsection{Eulerian-Lagrangian method}
\label{sec2p1}

The Eulerian-Lagrangian (EL) simulations presented here are based on the volume-filtered (VF) formulation \citep{anderson1967fluid,jackson2000dynamics,capecelatro2013euler}. 
In the EL formulation, the fluid phase is treated in the Eulerian frame as a continuum, while the particle phase is treated in the Lagrangian frame, with discrete particles tracked individually.

The volume-filtered conservation 
equations govern the fluid (carrier) phase in the semi-dilute regime (low particle volume fraction) as
\begin{subeqnarray}
\boldsymbol{\nabla} \cdot \textbf{u} =0~,\\
\rho_f\, \left(\frac{\partial \textbf{u}}{\partial t}+\textbf{u}\cdot\boldsymbol{\nabla}\textbf{u}\right) = -\boldsymbol{\nabla}p+\mu_f\, \nabla^2\textbf{u}+\textbf{F}_p~,
\label{eqn2p1}
\end{subeqnarray}
where $\textbf{u}$ is the fluid velocity, $p$ is the pressure, $\rho_f$ is the fluid density and $\mu_f$ is the fluid viscosity. 
The term $\textbf{F}_p$ represents momentum exchange from particles (dispersed phase) to fluid. For a semi-dilute concentration of particles ($\phi_p \ll 1$), $\textbf{F}_p$ can be expressed as
\begin{equation}
    \textbf{F}_p = -\phi_p\, \left. \boldsymbol{\nabla}\cdot \boldsymbol{\tau}\right|_p+\rho_p\, \phi_p\, \frac{(\textbf{v}-\left. \textbf{u}\right|_p)}{\tau_p}~,
    \label{eqn2p2}
\end{equation}
where, $\rho_p$ is the particle density, $\phi_p$ represents the volume fraction of particles in the fluid medium, $\tau_p = \rho_p\, d_p^2/(18\, \mu_f)$ is the relaxation time scale of the particle to the fluid acceleration and $d_p$ the particle diameter. The total fluid stress, denoted as $\boldsymbol{\tau}$, is determined from the combination of pressure and viscous stresses as $-p\textbf{I} + \, \mu_f\, (\boldsymbol{\nabla} \textbf{u}+\boldsymbol{\nabla}\textbf{u}^{\textrm{T}})$, where $\boldsymbol{\nabla} \textbf{u}$ represents the fluid velocity gradient, $\textbf{I}$ is the identity matrix and $(\cdot)^{\textrm{T}}$ is the transpose operator. The Eulerian particle velocity, $\textbf{v}$, is computed from Lagrangian particle velocities using the equation (\ref{eqn2p4}). The notation $\left.(\cdot) \right|_p$ specifies fluid properties evaluated at the locations of the particles. In equation (\ref{eqn2p2}), the first term represents the stress exerted by the undisturbed flow at the location of the particle. The next term accounts for stresses induced by the presence of particles, characterized by Stokes drag, for $Re_p \ll 1$, where $Re_p$ is the Reynolds number based on particle size. The Stokes drag must be evaluated using the slip velocity between the particle and the undisturbed fluid. In situations where the density ratio ($\rho_p/\rho_f$) is significantly higher (e.g., dusty gas flows, water droplets in the air), as in our study, Stokes drag dominates the momentum exchange.

The particles are tracked in a Lagrangian sense. Assuming point spherical particles in a $Re_p \ll 1$ flow regime, the dynamic equation for $\textit{i}^{th}$ particle is given by \citet{maxey1983equation} as 
\begin{equation}
    \frac{d \textbf{v}^i}{d t} = \frac{1}{\rho_p}\, \left. \boldsymbol{\nabla} \cdot \boldsymbol{\tau}\right|_p+\frac{\left. \textbf{u}\right|_p-\textbf{v}^i}{\tau_p}~, \textrm{with} \, \quad \frac{d \textbf{x}^i}{d t} = \textbf{v}^i~,
    \label{eqn2p3}
\end{equation}
where, $\textbf{x}^i$ and $\textbf{v}^i$ are the position and velocity of the $\textit{i}^{th}$ particle respectively. In this study, gravitational/sedimentation effects have been omitted to isolate and comprehend the distinctive impact of two-way coupling on instability. Also, we operate in the semi-dilute regime to avoid any potential particle-particle interactions such as collision. Also, as mentioned earlier, the density ratio ($\rho_p/\rho_f$) is kept large, so the added mass effect and Basset history effect are negligible. To evaluate the momentum feedback from the particle phase to the fluid phase ($\textbf{F}_p$), one needs to evaluate the instantaneous particle volume fraction and Eulerian particle velocity field. At a location $\textbf{r}$, these are obtained from the corresponding instantaneous Lagrangian quantities using
\begin{subeqnarray}
    \phi_p(\textbf{r}) = \sum_{i = 1}^{N}V_p\, g(\lVert \textbf{r}-\textbf{x}^i\rVert)~,\\
        \phi_p(\textbf{r}) \, \textbf{v}(\textbf{r}) = \sum_{i = 1}^{N}\textbf{v}^i\,V_p\, g(\lVert \textbf{r}-\textbf{x}^i\rVert)~,
        \label{eqn2p4}
\end{subeqnarray}
where $V_p = \pi\, d_p^3/6$ is the particle volume, $g$ represents a Gaussian filter kernel of size $\delta_f = 3\, \Delta x$, where $\Delta x$ is the grid spacing \citep{capecelatro2013euler}. In the VF method, the Gaussian kernel smooths out/regularizes the fluctuations in momentum feedback, thereby preventing any convergence issues during simulation. 
The VF model was recently applied by the authors successfully in particle-laden vortical flows \citep{shuai2022accelerated,shuai2022instability,shuai2024merger}. Readers interested in further details about the numerical approach may refer to them.


A scaling analysis of the drag force in equation (\ref{eqn2p2}) reveals the coupling strength of feedback force from particle phase to fluid phase is governed by the non-dimensional number $M = \rho_p\, \langle \phi_p \rangle/\rho_f$ - the mass loading (or mass fraction), where $\langle \phi_p \rangle$ is the average volume fraction of particles. If the particle field is dilute and the mass loading is negligible, the feedback force can be neglected, and the one-way coupled simulations can be used to describe the evolution of the particulate flow. However, when the density ratio  ($\rho_p/\rho_f$) is significant, as is the case here, the mass loading becomes $\textit{O}(1)$, which leads to significant feedback to the fluid phase from the particle phase, even if the particle phase is dilute \citep{kasbaoui2015preferential}. As we will see in the upcoming sections, this feedback is the source of instability in the present study, as the system considered here is stable under one-way coupling.

\subsection{Illustration of the instability}
\label{sec2p2}
To demonstrate the instability resulting from two-way coupling, we investigate an unbounded simple shear flow combined with a band of particles distributed in a top-hat manner (refer to the schematic in Figure \ref{fig1}). The fluid properties include a density of $\rho_f = 1.0 \,\textrm{Kg\, m}^{-3}$, viscosity of $\mu_f = 7.9\times10^{-5}\, \textrm{Kg\, m}^{-1}\,\textrm{s}^{-1}$, and a shear rate of $\Gamma = 10^{-2}\, \textrm{s}^{-1}$. The particles are mono-disperse with a diameter $d_p = 377 \, \mu \textrm{m}$ and density $\rho_p = 1000 \,\textrm{Kg\, m}^{-3}$, distributed uniformly within a region $\lvert y \rvert \leq h/2$. The band width is $h = 0.5\, \textrm{m}$, and the average volume fraction of particles within the band is $\langle \phi_p \rangle = 10^{-3}$. In terms of nondimensional numbers, this corresponds to a density ratio of $\rho_p/\rho_f = 1000$, Stokes number $St =\Gamma\, \tau_p= 10^{-3}$, and mass loading $M = (\rho_p/\rho_f)\, \langle \phi_p \rangle = 1$, which is significant. Here, $\tau_p = \rho_p\, d_p^2/(18\, \mu_f)$ is the particle relaxation time.

The numerical simulation is performed in a box of dimensions $L_x\times L_y\times L_z$, where $L_x = 4\pi \, h$, $L_y = 3\, L_x$, and $L_z = 3\, d_p$. To avoid unwanted diffusion effects, the Reynolds number based on the box width is thus set to $Re_{L_x} = \Gamma\, L_x^2\rho_f/\mu_f = 5000$, and we focus on inviscid instability. The flow needs to be periodic in the $x$ direction and unbounded in the $y$ direction. To achieve this within the simulation box, we apply regular periodic boundary conditions at the left and right boundaries (in the $x$ direction) and shear-periodic boundary conditions at the top and bottom boundaries (in the $y$ direction), accounting for the background shear flow \citep[see][]{kasbaoui2017algorithm}. By choosing a domain size that is three times larger in the $y$ direction, we ensure neighbouring periodic simulation boxes are well-separated and do not interfere with each other to influence the instability. We use the EL framework described above on a uniform Cartesian grid with resolution $h/\Delta x  \approx 42$, $N_x = 512$, $N_y = 3\, N_x$ and $N_z = 1$.
Here, we perform pseudo-two-dimensional simulations by considering only one grid point in the axial ($z$) direction with periodic boundary conditions applied over a thickness $\Delta z = 3 d_p$. This allows the definition of volumetric quantities such as particle volume and volume fraction. 

In addition to conducting two-way coupled simulations, we perform one-way coupled simulations, where the momentum exchange term (\ref{eqn2p2}) is neglected. The particles still evolve due to the momentum contribution from the fluid. However, the particles' feedback to the fluid phase is deliberately switched off. This allows for comparison between one-way and two-way coupled simulations and showcases the impact of particle feedback on the flow dynamics.
\begin{figure}
    \centering
    \includegraphics[width=1.0\linewidth]{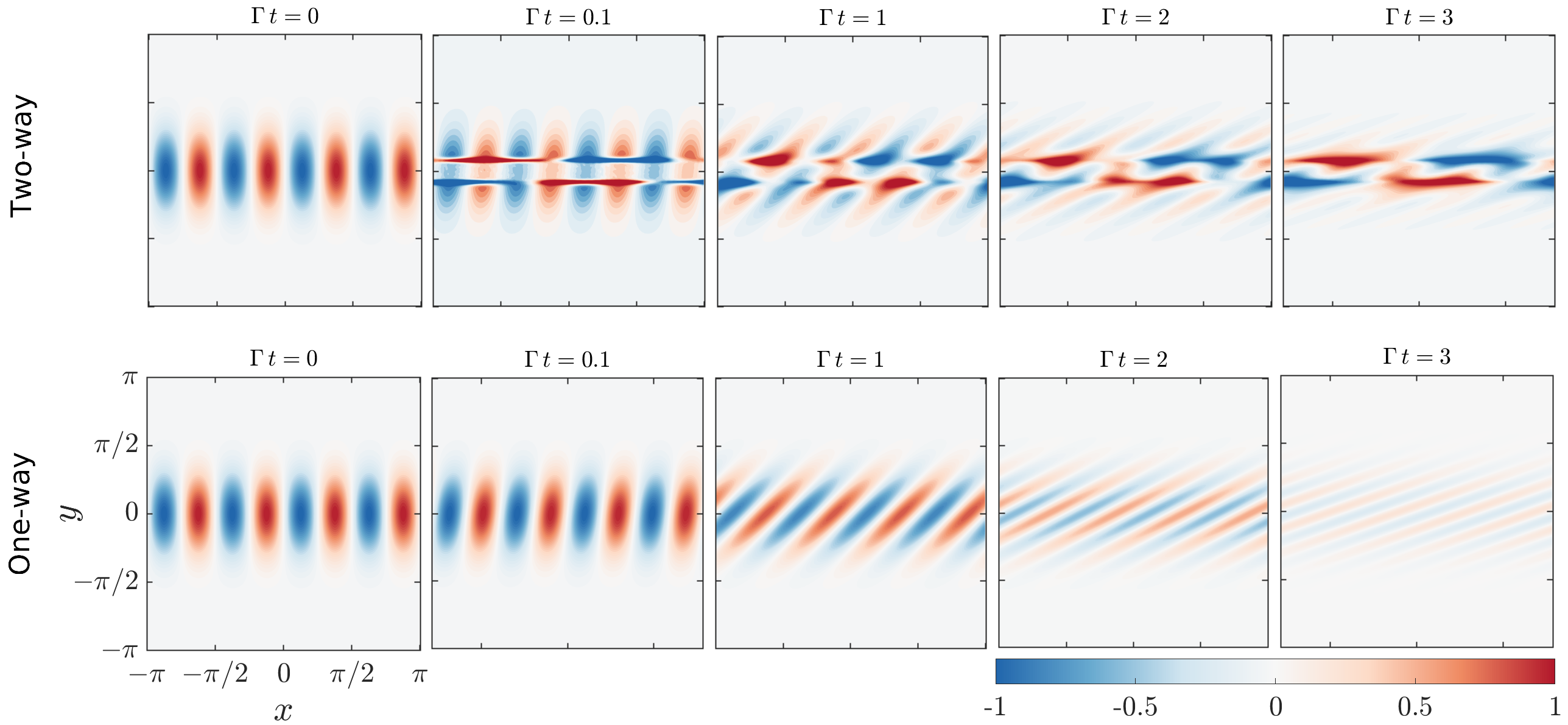}
    \caption{Isocontours of perturbation vorticity ($\tilde{q}_z$) normalized with the maximum initial perturbation vorticity for two-way coupling (top panel) and one-way coupling (bottom panel), shown for various non-dimensional times $\Gamma\, t = 0, 0.1, 1, 2$ and $3$. The corresponding parameters are set to $M=1$, $St = 10^{-3}$, $\epsilon = 10^{-2}$, and $\langle \phi_p \rangle = 10^{-3}$. 
    }
    \label{fig2}
\end{figure}
\begin{figure}
    \centering
    \includegraphics[width=1.0\linewidth]{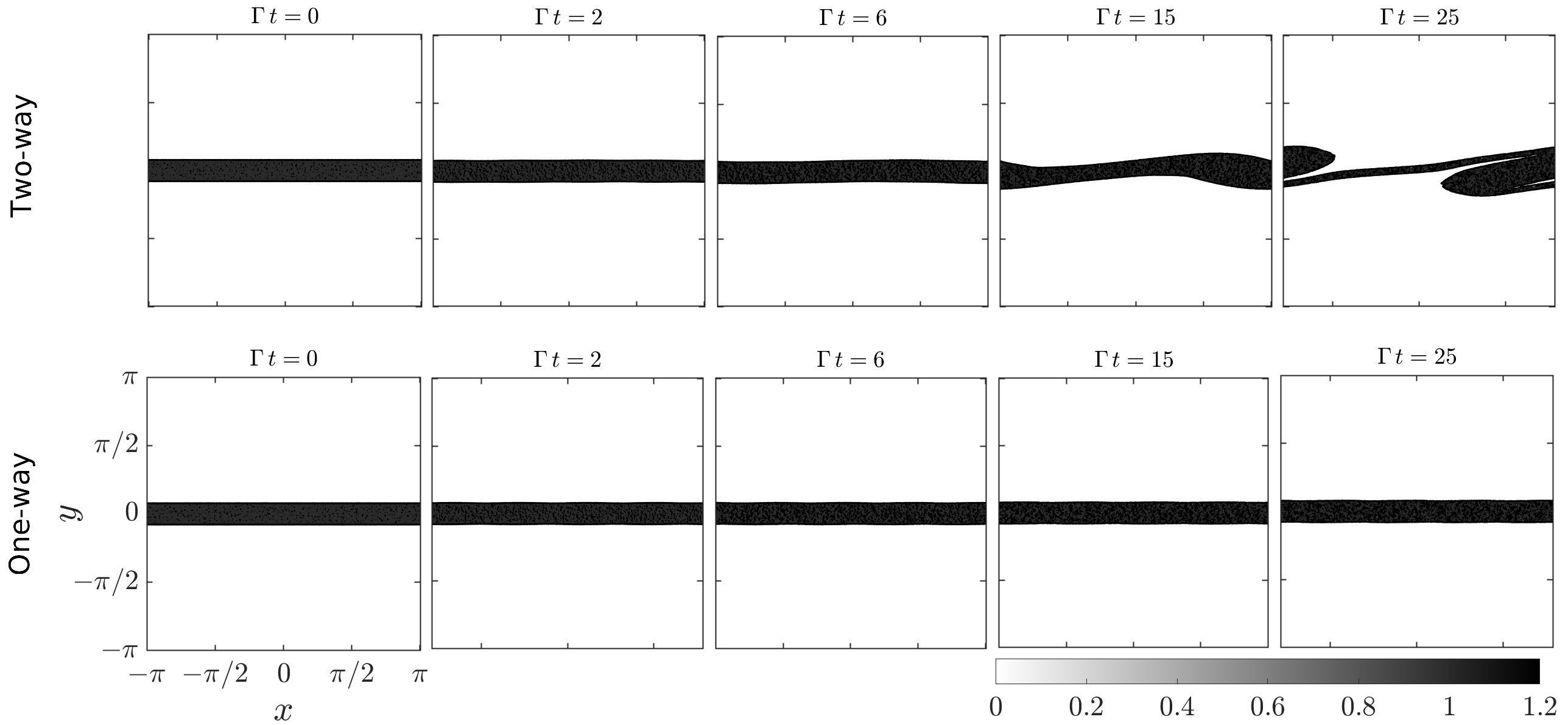}
    \caption{Isocontours of the non-dimensional particle number density ($\phi_p/\langle \phi_p \rangle$), corresponding to the simulation in figure \ref{fig2}, are depicted for various non-dimensional times 
    $\Gamma\, t = 0, 2, 6, 15$ and $ 25$. The longer duration illustrates the nonlinear evolution of instability in the two-way coupling case. }
    \label{fig3}
\end{figure}

The fluid velocity field is initially superimposed with a two-dimensional incompressible perturbation of the form $\tilde{u}_x = \epsilon\, \Gamma\, h\, e^{-y^2}\, (y/2)\,\sin 4\, x$,  $\tilde{u}_y = \epsilon\, \Gamma\, h\, e^{-y^2}\, \cos 4\, x$, with perturbation amplitude $\epsilon = 10^{-2}$. The corresponding perturbation vorticity field is shown in figure \ref{fig2}, at $\Gamma \, t = 0$. The initial velocity of the particles is set equal to the local fluid velocity.

The evolution of the vorticity perturbation $\tilde{q}_z$ normalised by its initial maximum value is presented in figure \ref{fig2}. Successive snapshots are provided for various non-dimensional times $\Gamma\, t = 0, 0.1, 1, 2$, and $3$. The snapshots are confined to a domain size of $2\,\pi \times 2\,\pi$ for better visualisation, although the simulations employ a domain size of $2\,\pi \times 6\,\pi$, as mentioned earlier. In the case of one-way coupling (bottom panels), it is observed that the vorticity patches are sheared, tilted and stretched by the background flow, and the intensity of vorticity diminishes as time progresses. The downstream tilt of the vorticity perturbations and the related algebraic decay of associated perturbation energy by the Orr mechanism are described in detail in \cite{farrell1987developing,roy2014inviscid}.
 The perturbed flow had a periodic behaviour with a wavenumber in the x-direction of $k = 4 \, \textrm{m}^{-1}$, which remains unchanged as time advances. At later times, it can be seen that the perturbation field eventually decays.

Conversely, when considering two-way coupling (top panels), we observe significant evolution and amplification of the vorticity field. Initially, the perturbed mode with $k = 4\, \textrm{m}^{-1}$ disappears, giving way to a new mode with $k = 1\, \textrm{m}^{-1}$. These newly emerged structures, likely unstable eigenmodes, persist, and their corresponding vorticity field intensifies over time. As the simulation progresses, nonlinear interactions between successive vorticity patches become more pronounced, eventually leading to a transition into a strongly nonlinear regime (not shown here).

Particle dispersion is also significantly affected when two-way coupling is considered. Figure \ref{fig3} depicts the evolution of the particle volume fraction field scaled with the average initial volume fraction. When the particle feedback is neglected (bottom panels), the shear flow simply advects the particles. As the flow perturbations decay over time, they have minimal impact on particle transport, even after a significantly longer duration. Eventually, the disturbances die out, and the particle distribution resembles the initial band (see figure \ref{fig3}, bottom panel, at $\Gamma \, t = 0$ and $25$). In contrast, two-way coupling leads to growing flow perturbations, which in turn govern the dispersion of particles (top panels). Initially, the uniform particle band undergoes deformation due to flow disturbances characterized by a periodic mode of $k = 1\, \textrm{m}^{-1}$. As time progresses, the interplay between background shear flow and growing perturbations initiates nonlinear effects, resulting in particle clustering into lobes interconnected by a relatively slender filament, as can be seen in figure \ref{fig3}, top panel, at $\Gamma \, t=25$.

The simulations shown in figures \ref{fig2} and \ref{fig3} suggest that semi-dilute inertial particles, distributed non-uniformly in an unbounded simple shear flow, can induce hydrodynamic instability. This instability cannot be solely attributed to hydrodynamics since the simple shear flow in a single-phase flow is stable to infinitesimal perturbations \citep[see][]{drazin2004hydrodynamic}. Furthermore, it cannot be attributed to collisional effects, as the simulation neglected particle-particle interactions. Instead, the instability must arise from the two-way momentum exchange between the two phases, as confirmed by the absence of instability when the two-way coupling term is deactivated in the simulation. Previous studies have shown that uniformly distributed particles in a simple shear flow, even with two-way coupling, do not exhibit modal instability but can demonstrate only a non-modal instability if gravitational effects are considered \citep{kasbaoui2015preferential}. However, in this study, we observe the emergence of a new type of instability resulting from the interaction between the simple shear flow and a non-uniformly distributed particle field in the absence of gravitational settling. In the subsequent sections, we will demonstrate that this new type of instability is modal and explain its generation mechanism. The following section will provide a mechanistic explanation of the instability through wave interactions.

\section{Mechanism of instability: A dusty Taylor-Caulfield instability}
\label{sec3}
In the previous section, we observed that instability arises from the two-way coupling between the particle and fluid phases, driven by the finite inertia of the particles. Surprisingly, even weakly inertial particles ($St = 10^{-3}$) triggered the instability. In this section, we delve into the instability mechanism in the small particle inertia limit while still considering the particle-fluid coupling. We employ the concept of wave interaction to elucidate this instability mechanism. In the weak particle inertia limit, we will demonstrate that our particle-laden system resembles a stratified fluid system. The fluid exhibits an effective modified density, which is stratified due to the particle concentration gradient. This density stratification creates edge waves, and their interaction leads to instability, similar to the case of the Taylor-Caulfield instability \citep{taylor1931effect,caulfield1994multiple,balmforth2012dynamics}. However, there is a caveat: the wave generation mechanism differs slightly here, which we will address as we proceed.

In the limit of weak particle inertia, following \cite{ferry2001fast,rani2003evaluation}, the feedback force $\textbf{F}_p$ in equation (\ref{eqn2p2}) for dusty-gas system can be approximated as $\textbf{F}_p = -\rho_p\,\phi_p\, \textbf{a}+\textit{O}(\tau_p)$, where $\textbf{a} = \left( \partial \textbf{u}/\partial t+\textbf{u}\cdot\boldsymbol{\nabla}\textbf{u}\right)$ represents the fluid acceleration term. Substituting it back into equations (\ref{eqn2p1}b) yields a simplified form representing a fluid with modified density (in the inviscid limit) as
\begin{equation}
\rho\, \left(\frac{\partial \textbf{u}}{\partial t}+\textbf{u}\cdot\boldsymbol{\nabla}\textbf{u}\right) = -\boldsymbol{\nabla}p~,
\label{eqn3p1}
\end{equation}
where $\rho = \rho_f+\rho_p\, \phi_p$ represents an effective fluid density due to the suspended particles. Thus, a spatial inhomogeneity in the particle concentration ($\phi_p$) can result in a variation in the effective density of this composite fluid even if $\rho_f$ is a constant. Additionally, the conservation of the number of particles yields
\begin{equation}
    \frac{\partial \rho}{\partial t}+\textbf{u} \cdot \boldsymbol{\nabla}\rho  = 0~.
    \label{eqn3p2}
\end{equation}
Together, equations (\ref{eqn2p1}a), (\ref{eqn3p1}) and (\ref{eqn3p2}) resemble a stratified incompressible fluid system and is known as the single-fluid continuum model describing a particle-laden system. Taking the curl of equation (\ref{eqn3p1}) after scaling it with $\rho$ gives the evolution equation for the flow vorticity ($\textbf{q} = \boldsymbol{\nabla} \times \textbf{u}$) as (for a two-dimensional case)
\begin{equation}
    \left(\frac{\partial \textbf{q}}{\partial t}+\textbf{u}\cdot\boldsymbol{\nabla}\textbf{q}\right) = \textbf{a} \times\frac{\boldsymbol{\nabla} \rho}{\rho}~,
    \label{eqn3p3}
\end{equation}
where, we have used the relation between $\textbf{a}$ and $\boldsymbol{\nabla}p$ from equation (\ref{eqn3p1}) here for further simplification. Equation (\ref{eqn3p3}) indicates that vorticity can arise in the system due to the misalignment between fluid acceleration and the gradient in particle concentration due to the baroclinic source term $\textbf{a} \times \boldsymbol{\nabla}\rho$. In the subsequent paragraphs, we will demonstrate the precise way by which vorticity is generated and how it gives rise to propagating waves that can interact and lead to instability.

The base state flow may inherently possess a vorticity field. However, the additional vorticity disturbance created would be responsible for generating waves. To analyze this, we need to consider the evolution equation for the disturbance vorticity field. Without loss of generality, let's consider a general two-dimensional system in the $x$--$y$ plane with uni-directional flow in the horizontal direction and vertically stratified particle distribution. We can decompose the relevant quantities into base state and disturbance (denoted by $\tilde{()}$) parts as $\textbf{u} = U(y)\, \hat{\textbf{x}} + \tilde{\textbf{u}}(x,y)$ and $\rho = \rho_b(y) + \tilde{\rho}(x,y)$. The vorticity field $\textbf{q}=q_z\, \hat{\textbf{z}}$ will be along the $\hat{\textbf{z}}$ direction and can be decomposed as $q_z = Q_z(y) + \tilde{q}_z(x,y)$, where the base state vorticity is $Q_z(y) = -U'(y)$. Substituting these expressions into equation (\ref{eqn3p3}) and considering the leading disturbance terms, we obtain the linearized evolution equation for the disturbance vorticity field as
\begin{equation}
    \frac{\mathcal{D} \tilde{q}_z}{\mathcal{D} t} = \frac{\rho_b'}{\rho_b}\, \frac{\mathcal{D} \tilde{u}_x}{\mathcal{D} t}+\frac{(\rho_b\, U')'}{\rho_b}\, \tilde{u}_y~,
    \label{eqn3p4}
\end{equation}
where the operator $\mathcal{D}/\mathcal{D} t = \partial/\partial t+U(y)\, \partial/\partial x$ represents the linearized material derivative, and $()'$ represents the operation $d/dy$. Utilizing the relationship between the vertical displacement field $\tilde{\eta}$ and the vertical velocity field $\tilde{u}_y = \mathcal{D}\tilde\eta/\mathcal{D} t$, we can rewrite equation (\ref{eqn3p4}) as
\begin{equation}
    \frac{\mathcal{D} }{\mathcal{D} t} \left(\tilde{q}_z-\tilde{u}_x\, \frac{\rho_b'}{\rho_b}-\tilde{\eta}\, \frac{(\rho_b\, U')'}{\rho_b}\right) = 0~,
    \label{eqn3p5}
\end{equation}
i.e., the quantity inside the bracket is materially conserved. In other words, the disturbance vorticity $\tilde{q}_z$ is related to the horizontal disturbance velocity $\tilde{u}_x$ and the vertical displacement $\tilde{\eta}$ as
\begin{equation}
    \tilde{q}_z=\tilde{u}_x\, \frac{\rho_b'}{\rho_b}+\tilde{\eta}\, \frac{(\rho_b\, U')'}{\rho_b}+\textrm{constant}~.
    \label{eqn3p6}
\end{equation}
\begin{figure}
\centerline{\includegraphics[width=0.85\linewidth]{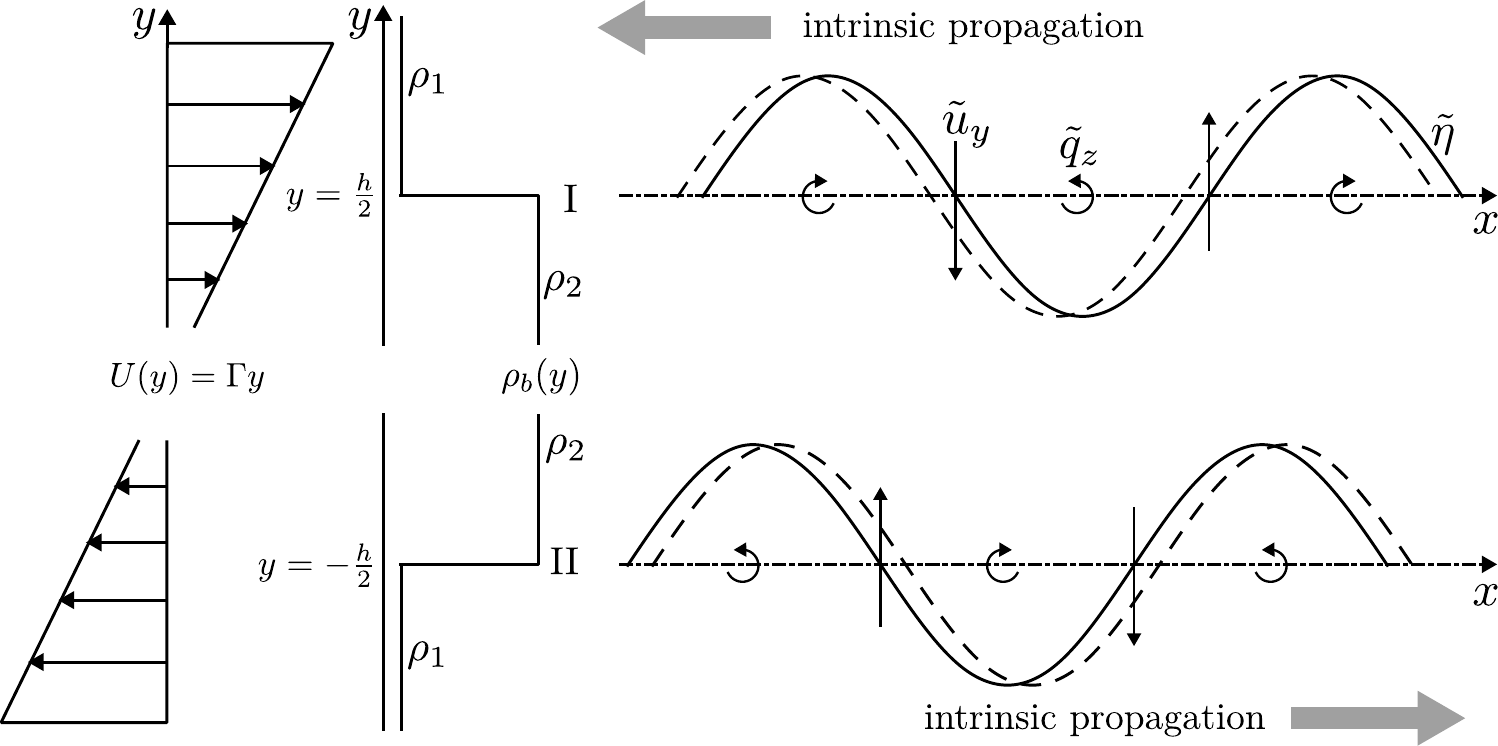}}
  \caption{Schematic illustrating the background simple shear flow, density jumps at the two interface locations labelled $\textrm{I}$ and $\textrm{II}$, and the corresponding interface disturbance fields. The initial interface displacement field ($\tilde{\eta}$) is depicted as a continuous sinusoidal curve, while its later stage is shown as a dashed curve, indicating its intrinsic propagation direction. The perturbation vorticity field ($\tilde{q}_z$) and perturbation vertical velocity fields ($\tilde{u}_y$) are also sinusoidal, with crests (troughs) represented by counter-clockwise (clockwise) and upward (downward) arrows, respectively.}
\label{fig4}
\end{figure}
The constant term represents a bulk vorticity in the background flow. Since we are primarily interested in the relative vorticity generation $\tilde{q}_z$ with respect to this bulk vorticity, we can set the constant term to zero without loss of generality. From this general formulation, let us consider our special case where the base state flow is a simple shear flow, i.e., $U = \Gamma\, y$, and the base state particle number density has a top-hat distribution. Without loss of generality, we assume $\Gamma > 0$. Since the particle concentration has sharp variations at locations $y = h/2$ and $y = -h/2$, the effective density $\rho$ of the fluid also varies sharply at these locations. The base state density $\rho_b(y)$ takes a constant value $\rho_1 = \rho_f$ outside the particle band and $\rho_2 = (\rho_f+\langle \phi_p \rangle\, \rho_p) > \rho_1$ within the particle band, as shown in schematic figure \ref{fig4}. These density jumps cause the locations of the jumps ($y = \pm h/2$) to act like interfaces separating different density fluids, marked as $\textrm{I}$ and $\textrm{II}$ in the figure \ref{fig4}. To begin with, we consider each of these interfaces in isolation and demonstrate that they support propagating waves due to disturbances.

Consider the interface-$\textrm{I}$ at $y = h/2$, where the effective density drops from $\rho_2$ to $\rho_1$ (as we move along the positive $y$ direction). As a result, $\rho_b'(y) = -\Delta\rho\, \delta(y-h/2) < 0$, where $\Delta\rho = (\rho_2-\rho_1)>0$ and $\delta(\cdot)$ represents a Dirac delta function. Let's assume the interface is perturbed sinusoidally ($\tilde{\eta}$) as shown in the figure \ref{fig4}. From equation (\ref{eqn3p6}), we can deduce that a disturbance vorticity field is generated, given by $\tilde{q}_z=( \tilde{u}_x+\Gamma\, \tilde{\eta})\, \rho_b'/\rho_b$. Generally, the associated horizontal perturbation velocity $\tilde{u}_x$ will have a discontinuity at the interface, which changes sign once we cross the interface. Physically, for a wave to be supported at the interface, there can be no self-induced $\tilde{u}_x$ for a wave-like solution at the interface. Thus, $\tilde{u}_x = 0$ at the interface. Then the generated vorticity disturbance is directly proportional to the interface perturbation as $\tilde{q}_z=\Gamma\, \tilde{\eta}\, \rho_b'/\rho_b$. Since $\Gamma > 0$, $\rho_b > 0$, and as we saw earlier $\rho_b' < 0$ at the interface-$\textrm{I}$, the generated vorticity disturbance will be out of phase with the interface displacement field. The maximum $\tilde{q}_z$ (counter-clockwise) occurs at the troughs of the $\tilde{\eta}$ field, and the minimum $\tilde{q}_z$ (clockwise) occurs at the crests of the interface perturbation, as can be seen in the figure \ref{fig4} (the crests and troughs of the generated vorticity disturbances are shown as circular arrows). Since the density gradient is localised at the interface, the resulting vorticity disturbance would also be localised at the interface as a Dirac delta function with sinusoidal variation in the flow direction. The generated disturbance vorticity field induces a vertical flow velocity field $\tilde{u}_y$, which has a maximum at the positive sloping node and minimum at the negative sloping node of the $\tilde{\eta}$.The vertical velocity field lags behind the interface displacement by $90^\circ$. The crests and troughs of the $\tilde{u}_y$ field are represented as upward and downward vertical arrows in the figure \ref{fig4}. The $90^\circ$ phase difference serves as the perfect criterion for the propagation of the interface displacement as a wave. Upon visual inspection of figure \ref{fig4}, one can intuitively deduce that the upward $\tilde{u}_y$ at the positively sloping $\tilde{\eta}$ node and similarly the downward $\tilde{u}_y$ at the negatively sloping $\tilde{\eta}$ node would result in an intrinsic propagation of the wave leftward relative to the base state flow at the interface location. Also, the $90^\circ$ phase difference ensures that the wave would not experience any amplification or damping but only undergo propagation \citep[see][]{carpenter2011instability}.

This is similar to the generation of a propagating edge wave in a background shear flow where there is a jump in vorticity/shear rate present \citep[see][]{valis2006atmospheric}. A generalised edge wave in a stratified fluid, which has both jumps in shear rate (from $\Gamma_2$ to $\Gamma_1$) and density (from $\rho_2$ to $\rho_1$) across an interface, should have a phase speed $c$ of the form
\begin{equation}
    c = U_\textrm{interface}+\frac{(\Gamma_1\,\rho_1-\Gamma_2\,\rho_2)}{k\, (\rho_1+\rho_2)}~.
    \label{eqn3p7}
\end{equation}
Here, we don't have a jump in shear rate (i.e., $\Gamma_1 = \Gamma_2 = \Gamma$), but only a jump in density. Thus, the resulting wave at the interface-$\textrm{I}$ would propagate with a phase speed of $c_\textrm{I} = \Gamma\, h/2 - \Gamma\, At/k$, where $k$ is the spatial wavenumber in the $x$ direction of the disturbance, and we used the definition of the Atwood number $At = (\rho_2-\rho_1)/(\rho_2+\rho_1)$, a non-dimensional number which frequently appears in the instability of density stratified flows. In non-dimensional terms (as one chooses the length scale to be $h$ and the time scale to be $\Gamma^{-1}$), the phase speed is $c_\textrm{I}^* = c_\textrm{I}/(\Gamma\, h) = 1/2 - At/k^*$, or the dispersion relation $c_\textrm{I}^*\, k^* = k^*/2 - At$, which we will formally obtain as a large $k^* = k\, h$ (i.e., large separation $h$) asymptotic expression of the dispersion relation of the full system in \S \ref{sec4p2}. Here, $()^*$ represents non-dimensional quantities.

Similarly, another propagating wave would be generated at interface-II in isolation. At this interface, since the particle concentration varies such that $\rho_b' = \Delta\rho\, \delta(y+h/2) > 0$, the vorticity disturbance generated would be localised at the interface and in phase with $\tilde{\eta}$. As a result, the vertical velocity disturbance field thus produced would lead $\tilde{\eta}$ by $90^\circ$, and thus, the intrinsic propagation of the wave would be rightward relative to the base state flow at the interface location, as shown in figure \ref{fig4}. The propagation speed would be $c_\textrm{II} = -\Gamma\, h/2 + \Gamma\, At/k$ or $c_\textrm{II}^* = -1/2 + At/k^*$.

Now that we have established that there would be a leftward propagating wave at interface-$\textrm{I}$ and a rightward propagating wave at interface-$\textrm{II}$ when they are in isolation or equivalently when the interfaces are far separated (i.e., $h \rightarrow \infty$ or $k^* \gg 1$), let us consider the scenario as the separation $h$ between the interfaces is reduced. One can expect the interfacial waves to perceive other's presence and interact as they come closer. This is because, although the interfacial vorticity fields are localised Dirac delta functions, the vertical velocity fields typically exhibit exponential decay (wave evanescence) away from the interface location (e.g., $\exp\left({-k\lvert y-h/2\rvert}\right)$ as we see in \S \ref{sec4p2} as the eigenfunction associated with $\tilde{u}_y$). As the waves approach each other, they could interact. Still, for modal instability to occur, two essential conditions must be satisfied \citep[see][]{carpenter2011instability}: (i) the phase speeds of the waves should be stationary with respect to each other (`phase-locking') and (ii) the relative phase of the waves should lead to the mutual growth of the interface disturbances. As the interfaces approach each other, the individual wave speeds also approach one another and are expected to become equal (and equal to zero, i.e., $c_\textrm{I} = c_\textrm{II} = 0$) for $k\, h = k^* = 2\, At$. However, along with this natural convergence, interactions play a role such that the waves become stationary relative to each other at a slightly larger separation (or non-dimensional wavenumber $k^*_\textrm{cutoff} > 2\, At$). The leftward propagating wave at interface-\textrm{I} can slow down the rightward propagating wave at interface-\textrm{II} and vice versa (for $\Gamma >0$), thus achieving phase-locking at a larger separation. Once the phase-locking is achieved, it can be maintained even though the isolated phase speeds are generally unequal (except at $k^* = 2\, At$) due to mutual interaction. Adjustments in the phase difference between the waves maintain an induced phase speed that exactly cancels the intrinsic propagation speed. Therefore, condition (i) would be satisfied for $0 < k^* < k^*_\textrm{cutoff}$ through adjustments in the relative phase difference of the waves.

Similar to the propagation criterion, upon visual inspection of figure \ref{fig4}, one can deduce that an upward $\tilde{u}_y$ component at the crests of $\tilde{\eta}$ and equivalently a downward $\tilde{u}_y$ component at the troughs of $\tilde{\eta}$ will lead to the growth of the wave. As the isolated waves become phase-locked, the interaction occurs in such a way that the $\tilde{u}_y$ field of one wave at an interface and the $\tilde{\eta}$ field of the other interface satisfy this criterion, and vice versa. Thus, by condition (ii), the phase-locked waves mutually trigger amplification and arrest propagation, leading to instability through wave interaction for any smaller separation of the interfaces (or non-dimensional wavenumbers in the range $0 < k^* < k^*_\textrm{cutoff}$). This intricate interaction between the phase-locked waves is fundamental to understanding the mechanism behind the generation of modal instability in such dust-stratified fluid systems.

We will now address the earlier caveat, highlighting an important distinction from the classical Taylor-Caulfield instability. In the Taylor-Caulfield instability, interfacial waves are generated due to a Boussinesq-type baroclinic torque driven by buoyancy/gravity effects. This occurs in a system with an overall stable stratification ($\rho_1 < \rho_2 < \rho_3$), where $\rho_1$ is at the top, $\rho_2$ is at the middle and $\rho_3$ is at the bottom layers. A pair of stable surface gravity waves are generated at each interface, separating these layers due to the Boussinesq baroclinic effects. One of them propagates leftward while the other propagates rightward; together, they interact to produce a stationary wave which is dampening at each interface. However, when the interfaces are brought closer, in a background shear flow ($\Gamma > 0$), the interaction between the leftward propagating wave at the upper interface and the rightward propagating wave at the lower interface leads to instability. 

However, in our system, no gravitational/buoyancy effects are present. Therefore, no restoring mechanism could generate propagating waves from Boussinesq baroclinic effects. Instead, the oft-neglected non-Boussinesq baroclinic effects are responsible for wave generation, acting as a different type of restoring mechanism. The waves generated closely resemble edge waves or vorticity waves rather than surface gravity waves, as only one stable propagating wave is generated at each interface. Similar to the Taylor-Caulfield instability, instability in our system requires a leftward propagating wave at the top interface and a rightward propagating wave at the bottom interface (noting that this preference is due to $\Gamma > 0$ and would be reversed if $\Gamma < 0$). Thus, for instability to occur, $\rho_3$ must be smaller than $\rho_2$, which we have assumed to be the same as $\rho_1$. However, if the system were configured with $\rho_3 = \rho_1 > \rho_2$, the direction of waves at each interface would be reversed. In such a scenario, the interaction would only amplify their propagation speed, suppressing phase-locking and avoiding any chance of instability. Similarly, if $\rho_3>\rho_2>\rho_1$ ($\rho_3<\rho_2<\rho_1$), then there would be leftward (rightward) propagation of waves at both the interfaces, whose interaction also can not produce instability. With the evidence of instability from EL simulations and an explanation of the mechanism involving interacting edge waves, we now proceed towards a quantitative evaluation of the growth rate of the instability using a linear stability analysis.
\section{Linear stability analysis (LSA)}
\label{sec4}
In this section, we employ an analytical approach, complemented by numerical assistance, to investigate the system more formally, illustrating the existence and criteria and quantifying the growth rate of the instability. To achieve this, we incorporate a continuum description of the particle phase, outlined below.
\subsection{Two-fluid model}
\label{sec4p1}
We adopt the Eulerian-Eulerian description of particle-laden flow, as outlined by previous works such as \citep{saffman1962stability,marble1970dynamics,druzhinin1995two,jackson2000dynamics,kasbaoui2015preferential}. The dispersed phase, or particle phase, is characterized by a set of conservation equations inspired from the kinetic theory of gases, employing a mono-kinetic particle-velocity distribution function. According to the method, the particle phase is considered a continuum fluid of zero pressure, valid for very dilute suspensions. The mass and momentum conservation equations for the fluid and particle phases in the semi-dilute regime, expressed in non-dimensional form, are 
\begin{subeqnarray}
\boldsymbol{\nabla} \cdot \textbf{u} =0~,\\
\frac{\partial n}{\partial t}+\boldsymbol{\nabla}\cdot(n\, \textbf{v}) = 0~,\\
\frac{\partial \textbf{u}}{\partial t}+\textbf{u}\cdot\boldsymbol{\nabla}\textbf{u} = -\boldsymbol{\nabla}p+\frac{1}{Re}\, \nabla^2\textbf{u}+M\, \frac{ n\, (\textbf{v}-\textbf{u})}{St}~,\\
\frac{\partial (n\,\textbf{v})}{\partial t}+\boldsymbol{\nabla} \cdot \left(n\, \textbf{v}\,\textbf{v}\right) =  \frac{n\, (\textbf{u}-\textbf{v})}{St}~,
\label{eqn4p1}
\end{subeqnarray}
where $\textbf{u}$ represents the fluid velocity, $\textbf{v}$ denotes the particle velocity, and $n = \phi_p/\langle \phi_p \rangle$ is the normalized particle number density, all depicted as fields. Here onwards, we drop $()^*$ from non-dimensional quantities as we only deal with them unless specified otherwise. For non-dimensionalization, the length and time scales are set to the initial width of the particle band ($h$) and the inverse shear rate ($\Gamma^{-1}$), respectively.
The dimensional number density $\phi_p/V_p$ is normalized by $\langle \phi_p \rangle/V_p$ to obtain a non-dimensional measure for the particle concentration field denoted as $n = \phi_p /\langle \phi_p \rangle$, where $V_p$ represents the particle volume. The non-dimensional numbers include the Reynolds number ($Re=\Gamma\, h^2/\nu$), which quantifies fluid inertia, and the Stokes number ($St = \Gamma\, \tau_p$), characterizing particle inertia. We have not accounted for particle interactions and inertial lift forces in our current formulation, although these effects could alter particle trajectories significantly. Both hydrodynamic and non-hydrodynamic interactions among dust particles can influence particle trajectories, affecting phenomena like coagulation and deposition (see \cite{patra2022collision} and references therein). In this study, we focus on the highly dilute limit and neglect the role of any interactions. The non-zero slip velocity a particle has with its background local shear flow causes it to experience an inertial lift force, as derived by \citet{saffman1965lift} for a simple shear flow scenario. The magnitude of the Saffman lift force depends on the particle Reynolds number based on the local shear rate. In our problem, this Reynolds number is small hence we disregard the effect of inertial lift.
\subsubsection{Linearization and normal mode analysis of the inviscid equations}
\label{sec4p1p1}
We disregard viscous effects in the analytical approach and focus solely on the inviscid scenario, where $Re \rightarrow \infty$, as we have seen that the mechanism of the instability is inherently inviscid. 
For stability analysis, we linearize the governing equations (\ref{eqn4p1}). The flow quantities are split into their base state, denoted by capital letters, and perturbation, denoted with a tilde, such as $\textbf{u} = \textbf{U}+\tilde{\textbf{u}}$, $\textbf{v} = \textbf{V}+\tilde{\textbf{v}}$, $p = P+\tilde{p}$, and $n = N+\tilde{n}$. We consider a two-dimensional stability analysis in the $x$ -- $y$ plane. A general base state of the form $\textbf{U} =\textbf{V}= U(y)\, \hat{\textbf{x}}$, and $N = N(y)$ satisfies the governing equations (\ref{eqn4p1}). That is, it is assumed that there is no slip between the base state particle and fluid velocities. Perturbing the base state with two-dimensional infinitesimal disturbances of the form $\tilde{\textbf{u}}(x,y,t)$, $\tilde{\textbf{v}}(x,y,t)$, $\tilde{p}(x,y,t)$, and $\tilde{n}(x,y,t)$, we obtain the linearized version of the equations as 
\begin{subeqnarray}
    \boldsymbol{\nabla} \cdot \tilde{\textbf{u}} = 0~,\\
       \frac{\partial \tilde{n}}{\partial t} + U\, \frac{\partial \tilde{n}}{\partial x} + \tilde{v}_y\, \frac{d N}{d y} +N\, \boldsymbol{\nabla} \cdot \tilde{\textbf{v}} = 0~,\\
    \frac{\partial \tilde{\textbf{u}}}{\partial t} + U\, \frac{\partial \tilde{\textbf{u}}}{\partial x} + \tilde{u}_y\, \frac{d U}{d y}\, \hat{\textbf{x}} = -\boldsymbol{\nabla}\, \tilde{p}+M\, \frac{ N\, (\tilde{\textbf{v}}-\tilde{\textbf{u}})}{St}~,\\ 
  \frac{\partial (N\, \tilde{\textbf{v}})}{\partial t} + U\, \frac{\partial (N\, \tilde{\textbf{v}})}{\partial x} + N\, \tilde{v}_y\, \frac{d U}{d y}\, \hat{\textbf{x}} = \frac{ N\,(\tilde{\textbf{u}}-\tilde{\textbf{v}})}{St}~.
   \label{eqn4p2}
\end{subeqnarray}
Here 
$\tilde{\textbf{u}} = \tilde{u}_x\, \hat{\textbf{x}}+\tilde{u}_y\, \hat{\textbf{y}}$ and $\tilde{\textbf{v}} = \tilde{v}_x\, \hat{\textbf{x}}+\tilde{v}_y\, \hat{\textbf{y}}$, where $\hat{\textbf{x}}$ and $\hat{\textbf{y}}$ are the unit vectors in $x$ and $y$ direction respectively. While it may appear that $N$ could be scaled out of all the terms in Equation (\ref{eqn4p2}d), it is retained because one must consider that in regions where there are no particles, $N = 0$, and $\tilde{\textbf{v}} = (N\,\tilde{\textbf{v}})/ N$ cannot be defined in these empty regions. We use two-dimensional disturbances, following Squire's theorem \citep{squire1933stability}, according to which the most unstable modes should be two-dimensional, which is valid for dusty flows as well \citep{sozza2022instability}.
To solve the system, we utilize a standard approach of normal mode analysis. Since the linearized equations (\ref{eqn4p2}) are homogeneous in $x$ and $t$, we assume a normal mode form for the solutions as $\left\{\tilde{\textbf{u}},\tilde{\textbf{v}},\tilde{p},\tilde{n}\right\} = \left\{\hat{\textbf{u}},\hat{\textbf{v}}, \hat{p}, \hat{n}\right\}(y)\, e^{i (k\, x-\omega\, t)}$, to obtain the eigenvalue system, similar to one in \cite{saffman1962stability} (see appendix \ref{appA}). Here $k$ is the non-dimensional wavenumber in the $x$ direction (scaled by $h$) and $\omega$ is the non-dimensional angular frequency (scaled by $\Gamma$). The complex wave speed is defined as $c = c_R+i\, c_I = \omega/k$. For temporal stability analysis (real-valued $k$), the real part $\Re(c)=c_R$ determines the phase speed of the wave propagation, and the imaginary part $\Im(\omega) = c_I\, k =  \sigma$ determines the growth rate. If $\Im(\omega)<0$, the mode is stable, while if $\Im(\omega)>0$, it is unstable. We need to solve the linear eigenvalue system equations (\ref{eqnA2}), seeking the eigenvalues $c$ (or $\omega$) and eigenfunctions $\left\{\hat{u}_x, \hat{u}_y, \hat{p},\hat{v}_x, \hat{v}_y, \hat{n}\right\}$, where $\hat{\textbf{u}}= \hat{u}_x \, \hat{\textbf{x}}+ \hat{u}_y\, \hat{\textbf{y}}$, and $\hat{\textbf{v}}= \hat{v}_x\, \hat{\textbf{x}}+ \hat{v}_y\, \hat{\textbf{y}}$.


Note that the particle velocity disturbance field influences the evolution of perturbation number density, although there is no feedback from number density perturbations to other quantities. To simplify the analysis, equations (\ref{eqn4p2}a,c \& d) can be combined to obtain a single equation by eliminating variables $\tilde{u}_x$, $\tilde{p}$, $\tilde{v}_x$, and $\tilde{v}_y$. Following the approach outlined by \cite{saffman1962stability,asmolov1998stability}, this can be achieved in the modal space to obtain a simplified nonlinear eigenvalue system involving only two variables $\hat{u}_y$ and $\hat{n}$ as
\begin{subeqnarray}
        \left[\mathcal{U}\, (D^2-k^2)-\mathcal{U}''\right]\hat{u}_y+M\, D\left[(U-c)\, \Lambda \, N' \, \hat{u}_y\right] = 0~,\\
        i\, k\, (U-c)\,\Lambda\, \hat{n}+D[N\, \Lambda^2] \, \hat{u}_y=0~,
    \label{eqn4p3}
\end{subeqnarray}
  where $\mathcal{U}= (U-c)\,(1+ M\, N\, \Lambda)$ denotes a modified base state velocity, and the damping factor $\Lambda = \left(1+i\, k\, St\, (U-c)\right)^{-1}$ describes the averaged frequency response of the particle phase to fluctuations in fluid velocity. We use $D(\cdot)$ or $(\cdot)'$ to represent the differential operator $\frac{d}{d y}$, where generally the former one is reserved for perturbation quantities and the latter one reserved for base state quantities. Equation (\ref{eqn4p3}a) represents a modified form of the Rayleigh equation in the context of instability of dusty rectilinear flows, which we will now refer to as the `dusty Rayleigh equation' or DRE for short. Note that the compact form of equation (\ref{eqn4p3}a) may misdirect the reader that the curvature of the number density field (i.e., $N''$) plays a role in the instability. However, the terms containing $N''$ exactly cancel out and do not have any role, which will be apparent from the form given in equation  (\ref{eqnA4}a).
A derivation of the equations (\ref{eqn4p3}) can be found in appendix \ref{appA}. These equations differ from those in \cite{saffman1962stability} in that the latter considers a uniform base state number density field, leading to the absence of terms containing gradients of particle concentration. Additionally, they consider a viscous case with finite Reynolds number effects. However, we consider an inviscid scenario and non-uniform particle distribution, resulting in DRE (\ref{eqn4p3}a) instead of the modified Orr-Sommerfeld equation obtained by \cite{saffman1962stability}.

\subsection{LSA for inertialess ($St = 0$) particles and the dusty Rayleigh criterion}
\label{sec4p2}
In this section, we show that the instability is present even in the vanishing particle inertia limit, so the origin of the instability is attributed to the mass loading, which quantifies the two-way coupling. In the inertialess limit, i.e. $St = 0$, the damping factor is $\Lambda = 1$, and the modified velocity becomes a linear mapping of the actual base flow as $\mathcal{U} = (U-c)\,(1+ M\, N)$. Thus equations (\ref{eqn4p3}) can be reduced to
\begin{subequations}
\begin{align}
     \rho\left[(U-c)\, (D^2-k^2)-U''\right]\hat{u}_y+\rho' \, \left[(U-c)\, D-U' \right]\hat{u}_y = 0~,\\
         i\, k\, (U-c)\, \hat{n}+ N'\, \hat{u}_y = 0~,
\end{align}
\label{eqn4p5}
\end{subequations}
where, we denote $\rho := 1+M\, N$. 
The term $\rho$ is the equivalent of a variable density, where its variation contributes to the inertial term, leading to the generation of non-Boussinesq baroclinic vorticity. Moreover, the same equations (\ref {eqn4p5}) can also be derived from linearization and normal-mode analysis of the single-fluid model \citep[see][]{ferry2001fast,rani2003evaluation}, as in the limit of $St \rightarrow 0$, the two-fluid model reduces to a stratified incompressible fluid system as we have seen in \S \ref{sec3}.\newline
\textit{The dusty Rayleigh criterion}: The classic Rayleigh equation results in a necessary criterion for instability known as the inflection point theorem, which states that the base state velocity profile should exhibit an inflection point within the domain, meaning that the quantity $U''$ should change sign somewhere within the flow domain. The equivalent criterion in the context of the equation (\ref{eqn4p5}a) is that the quantity $\Sigma = D[\rho\, U']$ should change sign within the flow domain. We have already seen the same term $\Sigma$ in the equation (\ref{eqn3p6}), which generates vorticity disturbances and, thus, instability. A derivation of the modified Rayleigh criterion/dusty Rayleigh criterion can be found in Appendix \ref{appB}. An equivalent statement for piecewise linear base state profiles is that the two interfaces must exhibit opposite-signed jumps in $(\rho\, U')$ across them.

We investigate the stability of a particle-laden simple shear flow, corresponding to the EL simulation in \S \ref{sec2p2}, where an unbounded simple shear flow interacts with particles localized within a band, as illustrated in figure \ref{fig1}. The respective non-dimensional form of the linear base state velocity profile is $U = y$, and the piecewise constant top-hat number density profile is
\begin{equation}
N(y) = 
\begin{cases} 
1, & \textrm{if } -\frac{1}{2} \leq y \leq \frac{1}{2}, \\
0, & \textrm{otherwise},
\end{cases}
\label{eqn4p6}
\end{equation}
where, as mentioned earlier, the inverse shear rate ($\Gamma^{-1}$) is used as the time scale, and the width of the band ($h$) is used as the length scale. In our case of a simple shear flow, the DRE stated above simplifies to a sign change for $N'$ in the flow domain, i.e., the gradient of the base state number density profile must exhibit a sign reversal for instability. In another sense, for the piecewise profile in equation (\ref{eqn4p6}), the instability criterion requires opposite-signed jumps in $\rho \, U'$ (or $N$ here) at the interfaces $y = \pm 1/2$, which is satisfied here. While the number density profile $N(y)$ defined in equation (\ref{eqn4p6}) is discontinuous, its form supports sharp gradients at the locations $y = \pm 1/2$, across which the sign flip occurs. It may be easier to visualize this sign flip if we smooth the number density profile, as we will see in equation (\ref{eqn4p10}) and figure \ref{fig7}. Overall, the base state meets the essential conditions for instability. However, instability is only assured once explicitly proven, as the dusty Rayleigh criterion serves as a necessary condition rather than a sufficient one. This section aims to analytically establish the presence of instability and pinpoint the parameter ranges where it manifests.

\begin{figure}
\centerline{\includegraphics[width=1.2\linewidth]{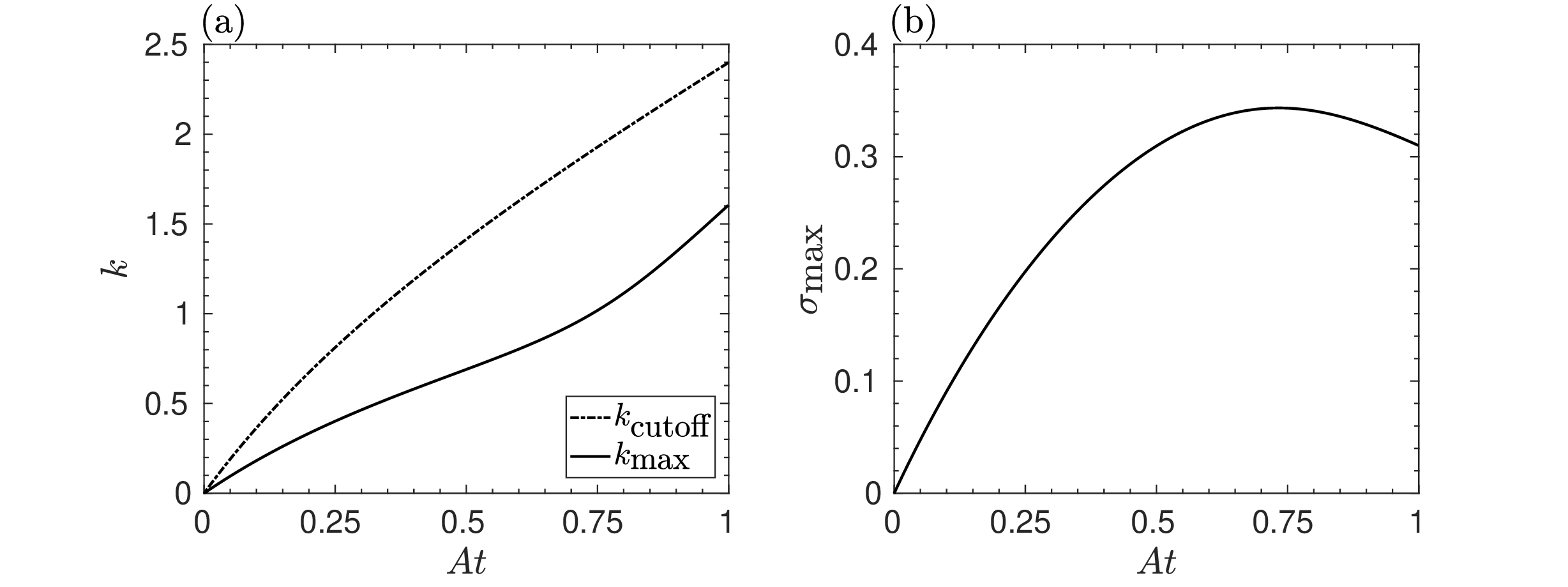}}
  \caption{(a) The variation of the cut-off wavenumber $ k_\textrm{cutoff} $, at which the instability disappears, and the optimum wavenumber $ k_\textrm{max} $, at which the maximum growth occurs, are plotted against the Atwood number for a top-hat number density profile. (b) The variation of the maximum growth rate $ \sigma_\textrm{max} $, corresponding to $ k_\textrm{max} $, with $ At $ is shown for the same number density profile.}
\label{fig5}
\end{figure}

For the linear velocity profile and top-hat number density profile, the stability equations (\ref{eqn4p5}a) reduce to $(D^2-k^2)\hat{u}_y = 0 $,
for all $y$ values except at the two locations where the number density profile has a discontinuity, i.e. at the interfaces $ y  = \pm 1/2$. The solution to this equation can be obtained analytically and is expressed piecewise for each layer as
 \begin{equation}
      \hat{u}_y = \begin{cases} 
      A_1\, e^{-k\, y}~, & y > 1/2 \\
      A_3\, e^{k y}+A_4\, e^{-k \, y}~, & \lvert y  \rvert <  1/2 \\
      A_2\, e^{k\, y}~, & y < -1/2
   \end{cases}
   \label{eqn4p7}
  \end{equation}
  where the unknown constants $A_1$ to $A_4$ need to be obtained from appropriate boundary conditions. Note that the decaying boundary condition for $\hat{u}_y$ at far-field is already accounted for in the solution, i.e. $\hat{u}_y(y \rightarrow \pm \infty) = 0$, assuming $k>0$. The kinematic criterion requires the continuity of $\hat{u}_y$ across the interfaces, ensuring the continuity of interface displacements ($\hat{\eta}$), as they are related through $i\, k\, (U-c)\, \hat{\eta} = \hat{u}_y$. Additionally, integrating equation (\ref{eqn4p5}a) across the discontinuous interfaces provides corresponding dynamic conditions. For detailed derivation, see the appendix \ref{appC}. Together, these conditions yield a system of four equations involving the unknowns $A_1$ to $A_4$ and $c$, as 
  \begin{equation}
     \begin{bmatrix}
     1 & 0 & - e^{k} & - 1 \\
     0 & 1 & -1 & -e^{k} \\
     \alpha_1 & 0 & \alpha_2\,e^{k} & -\alpha_2 \\
     0 & \alpha_3 & -\alpha_4 & \alpha_4\,e^{k} 
\end{bmatrix}\,     \begin{bmatrix}
     A_1 \\
      A_2 \\
      A_3 \\
     A_4 
\end{bmatrix}= \begin{bmatrix}
     0 \\
      0 \\
      0 \\
     0 
\end{bmatrix}~,
\label{eqn4p8}
\end{equation}
where $\alpha_1 = k\,\left(1/2-c\right)-M$, $\alpha_2 = k\,\left(1/2-c\right)\,(1+M)$, $\alpha_3=k\,\left(1/2+c\right)-M$ and $\alpha_4 = k\,\left(1/2+c\right)\,(1+M)$. For a nontrivial solution to the system of equations (\ref{eqn4p8}), the determinant of the coefficient matrix must be zero, leading to the dispersion relation as
\begin{equation}
    c = \pm\, \frac{1}{2\,k}\,\sqrt{\frac{(k-2\,At)^2-At^2\,(2+k)^2\, e^{-2\, k}}{1-At^2\, e^{-2\, k}}}~,
    \label{eqn4p9}
\end{equation}
where the Atwood number ($At=(\rho_\text{max}-\rho_\text{min})/(\rho_\text{max}+\rho_\text{min})$)
is related to the mass loading as $At = M/(M+2)$. Note that the eigenvalues $c$ appear in pairs, especially when $\Im(c) \neq 0$, they manifest as complex conjugates. Therefore, for instability to occur, $\Im(c)$ should not be zero. Equation (\ref{eqn4p9}) indicates the presence of a cut-off wavenumber ($ k_\textrm{cutoff}$), below which only the instability can occur, i.e., long-wave instabilities. This critical wavenumber can be determined as the solution of the transcendental equation $ M \left(2-k + (2+k)\, e^{- k}\right)-2\, k=0 $. The variation of $k_\textrm{cutoff}$ with $At$ is shown in the figure \ref{fig5}(a). For instance, for a mass loading of $M=1 \, (\textrm{or}\, At=1/3)$, the critical wavenumber can be obtained as $k_\textrm{cutoff} \approx 1.028$. 
The resulting unstable modes are stationary since the corresponding $\Re(c) = 0$. Conversely, the modes are neutrally stable propagating waves for shorter wavelengths, i.e. when $k > k_\textrm{cutoff}$. In the limit of small $k$ values, instability is always present but with a lower growth rate, as evidenced by the scaling $\omega \sim \pm i \, \sqrt{k}\, \sqrt{At/(1+At)}$.
\begin{figure}
\centerline{\includegraphics[width=\linewidth]{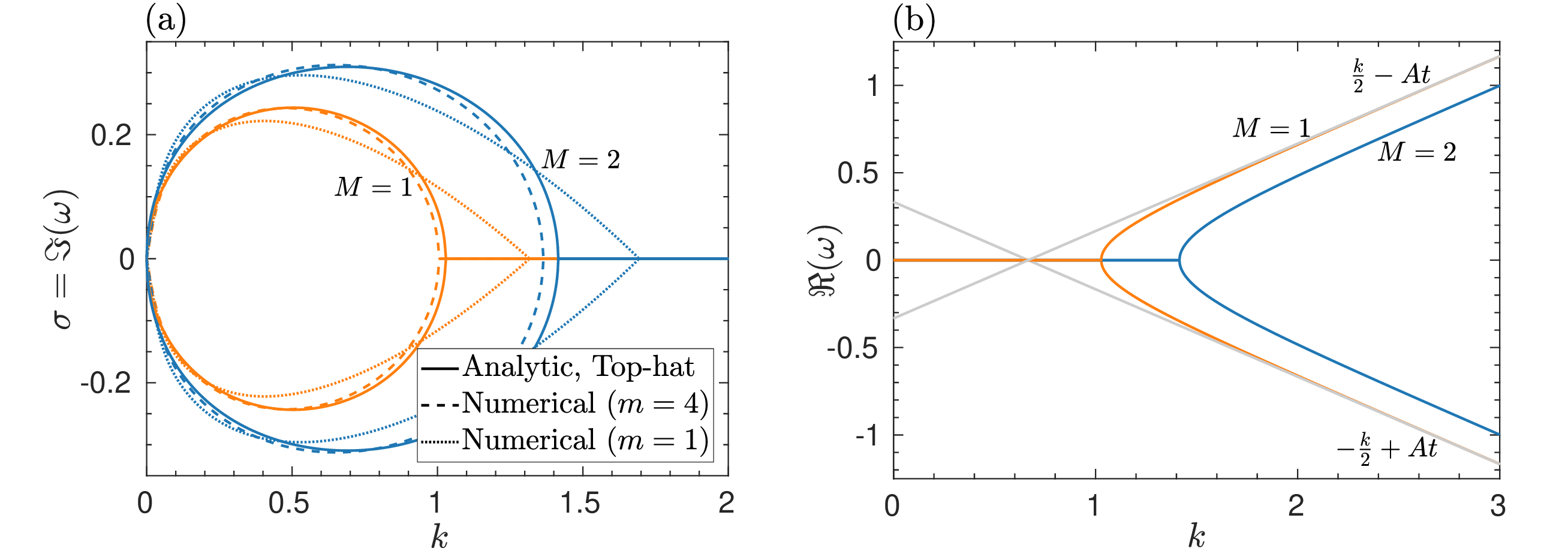}}
  \caption{Dispersion relation for the $St = 0$ case, for two different mass loadings ($M=1$ - orange and $M=2$ - blue), obtained analytically (for a top-hat number density profile) and numerically (for two different smooth number density profiles: $m=1$ and $m=4$): (a) Growth rate $\sigma$ (imaginary part of $\omega$) versus $k$, (b) Real part of $\omega$ versus $k$.  }
\label{fig6}
\end{figure}

Upon substituting the solution for $c$ in equation (\ref{eqn4p9}) back into equation (\ref{eqn4p8}), the amplitudes $A_1$ to $A_4$ can be determined, as detailed in appendix \ref{appC}. The dispersion relation (\ref{eqn4p9}) is plotted in figure \ref{fig6}, using continuous lines, for two different mass loading values. It can be observed from figure \ref{fig6}(a) that, for $k < k_\textrm{cutoff}$, the growth rate of the modes increases until reaching a maximum and then decreases for any given mass loading. Consequently, an optimal wavenumber ($k_\textrm{max}$) exists at which the maximum growth ($\sigma_\textrm{max}$) occurs. This optimal wavenumber can be obtained by solving the transcendental equation $e^{2\, k}\, (k-2\,At)+2\,At^2\, (1+2\, At)\,(1-At+ k)+e^{-2\, k}\,At^4\, (2+k)  = 0$, derived from equation (\ref{eqn4p9}) using the optimization criteria $d (c\, k)/d k = 0$. 
Figure \ref{fig5}(a) displays the variation of $k_\textrm{max}$ with $At$ while figure \ref{fig5}(b) shows the variation of $\sigma_\textrm{max}$ with $At$. For instance, for $M=1 \, (\textrm{or}\, At=1/3)$, the non-dimensional wavenumber corresponding to the maximum growth rate is $k_\textrm{max} \approx 0.504$, and the maximum growth rate is $\sigma_\textrm{max} = k_\textrm{max}\, \Im(c(k_\textrm{max})) \approx 0.244$. 

 From figure \ref{fig6}(b), it is evident that for $k > k_\textrm{cutoff}$, the waves propagate in opposite directions and attain a speed of $\omega \sim \pm (k/2 - At)$ as $k \rightarrow \infty$ (shown as asymptotes in the figure for $M=1$ in grey colour). This implies that the waves propagate without any interaction, as they are well separated by the distinct interfaces on which they exist. As we discussed the instability mechanism in \S \ref{sec3}, we obtained the same dispersion relation earlier for isolated waves at the interface. The natural phase speed matching of the isolated left and right propagating waves occurs at $k = 2\, At$, as indicated by the crossing of these asymptotic lines in the dispersion curve. However, in practice, this matching occurs at $k_\textrm{cutoff} > 2\, At$ due to their interaction, as seen from the figure.
\begin{figure}
\centerline{\includegraphics[width=1.2\linewidth]{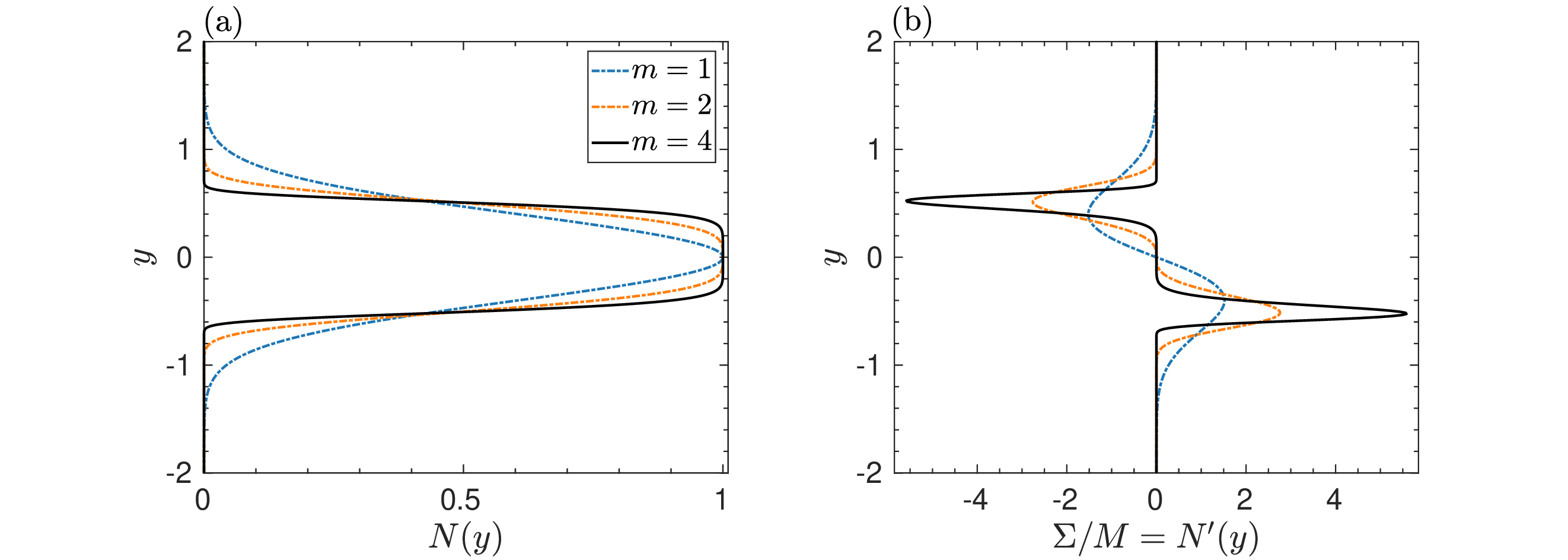}}
  \caption{(a) The smooth base state number density profile corresponds to the generalized Gaussian for various smoothness parameters $m$. (b) The corresponding parameter $ \Sigma $ (scaled with mass loading $ M $) in the background simple shear flow, which determines the stability, is plotted in the flow domain.}
\label{fig7}
\end{figure}

For the linear velocity profile and the top-hat number density profile, equation (\ref{eqn4p5}b) simplifies to $(y-c)\, \hat{n}=0$ everywhere except at the interfaces $y=\pm 1/2$. For the discrete modes (which corresponds to $y \neq c$), this equation yields $\hat{n} = 0$ everywhere except at the interfaces. Considering the interface discontinuity and conservation of the total number of particles, one finds $\int_{-\infty}^{\infty} \hat{n}\, dy = 0$, which implies $\hat{n} \propto \delta(y-1/2)-\delta(y+1/2)$.

To verify the analytical results, we numerically solved the linear eigenvalue differential equation (\ref{eqn4p5}a) using a standard Chebyshev spectral collocation method. To avoid issues arising from the discontinuity of the base state number density profile, we employed a smooth generalized Gaussian profile for the number density
  \begin{equation}
      N(y) = \exp \left\{-\left(\frac{y}{y^*}\,\right)^{2\, m} \right\}, \quad \textrm{with} \, \, \,  m \in \mathbb{N}
      \label{eqn4p10}
  \end{equation}
where, $y^* = \left( 2\,   \Gamma \left(1+1/(2\, m) \right) \right)^{-1}$, ensuring that the normalization yields $\int_{-\infty}^{\infty}N(y)\, dy = 1$. 
Here, $\Gamma(\cdot)$ denotes the Gamma function, and $m$ is the smoothness parameter. The number density profile exhibits maximum smoothness for small integer values of $m$, whereas for larger values of $m$, it develops sharp variations. As $m$ tends to infinity, the profile approaches the singular top-hat profile described in equation (\ref{eqn4p6}), as ensured by the normalisation. The plots in figure \ref{fig7} depicts the variation of the smooth base state number density profile and the corresponding stability-determining quantity $\Sigma $ for various smoothness parameters $ m $ within the flow domain. The Chebyshev collocation points ranging from $y \in [-1,1]$ are transformed to $y \in [-R,R]$, where $R$ is chosen to be sufficiently large to mimic infinity. To enhance the efficiency of the numerical method by capturing the sharp variations near the interfaces more effectively, we use the transformation as in \cite{govindarajan2004effect}, which ensures that more collocation points are clustered near $y = 0$. Since the base state profiles are now smooth, there is no need to apply interface conditions. Instead, only the far-field decay of the eigenfunctions needs to be enforced.

Figure \ref{fig6} illustrates the corresponding dispersion relation obtained for two different smooth profiles corresponding to the smoothness parameter $m=1$ (Gaussian) and $m=4$, depicted using dotted and dashed lines, respectively. Figure \ref{fig6}(a) shows that the smooth variation in the number density profile also causes instability. The trend of the dispersion curves remains the same, with a maximum growth rate at some optimum wavenumber. However,  the symmetry of the curve before and after this wavenumber is compromised for smooth profiles. Also, the cut-off wavenumber $k_\textrm{cutoff}$ at which the instability ceases to exist differs as the index $m$ changes. Note that, for the largest value of $m=4$, the numerical results compare well with the analytical results, with a better match expected for larger values of $m$. 

Figure \ref{fig6}(b) displays the variation of $\Re(\omega)$ with $k$, indicating the emergence of counterpropagating edge waves once the system is in the stable regime. The curves displayed are only for the analytical dispersion relation obtained for a top-hat number density profile. For smooth number density profiles, stable discrete modes are absent. The spectrum is purely continuous. The smooth variation of $ N' $ in this case, as opposed to the zero $ N' $ for a top-hat profile everywhere except the jump location, results in a logarithmic branch point at the critical layer ($ y=c $), as in the case of the continuous spectrum of inhomogeneous shear flows \citep{balmforth1995singular}.
The vorticity continuous spectrum eigenfunctions would be a combination of a delta-function singularity and a non-local Cauchy principal-value singularity (see \cite{van1955theory, case1959plasma}). However, the mechanistic picture of the interaction between edge waves offered earlier would persist despite the absence of discrete modes in the stable regime for smooth profiles. Interestingly, for smooth profiles with sufficiently ``steep" transition regions ($m\gg1$), a wave packet comprised of the continuous spectrum modes camouflages as a discrete mode known as a quasimode or a Landau pole \citep{schecter2000inviscid}. The quasimodes at each transition region can then interact and lead to instability, as seen previously in the computed growth rates for the smooth number density profile cases. 
\subsection{Effect finite particle inertia ($St \neq 0$)}
For any finite $St$, one must solve the equations (\ref{eqn4p3}) for simple shear flow ($U = y$). However, it is an analytically cumbersome task even for the top-hat number density profile in equation (\ref{eqn4p6}). The modified velocity $\mathcal{U}$ and the damping factor $\Lambda$ make this complication. As detailed in the appendix \ref{appD}, a small $St$ perturbation analysis can be performed to obtain asymptotic solutions. However, solving the equivalent linear eigenvalue system of equations (\ref{eqnA2}) numerically for a smooth number density profile would be more feasible. After eliminating $\hat{u}_x$ and $\hat{p}$ from equations (\ref{eqnA2}a,c \& d), the system can be reduced to the following form 
\begin{equation}
\renewcommand\arraystretch{1.5}
     \begin{bmatrix}
 l_1& \frac{M\, k}{St}\, l_2 & -\frac{i\, k^2\, M\, N}{St} & 0\\

-\frac{i}{k\, St}\, D  & l_3 & U' & 0\\
 -\frac{1}{St}  & 0 & l_3 & 0\\
 0 & i\, k\, N & l_2 & i\, k\, U
\end{bmatrix}\, \begin{bmatrix}
    \hat{u}_y\\
    \hat{v}_x\\
    \hat{v}_y\\
    \hat{n}
\end{bmatrix} = \omega\, \begin{bmatrix}

     (D^2-k^2) & 0 & 0 & 0 \\
     0 & i & 0 & 0 \\
     0 & 0 & i & 0 \\
     0 & 0 & 0 & i 
\end{bmatrix} \, \begin{bmatrix}
    \hat{u}_y\\
    \hat{v}_x\\
    \hat{v}_y\\
    \hat{n}
\end{bmatrix}~,
\label{eqn4p11}
\end{equation}
with the eigenvalues $\omega$ and the eigen functions $\left\{ \hat{u}_y,\hat{v}_x,\hat{v}_y,\hat{n} \right\}$. Here $l_1 = \left(k\, U-\frac{i\, M\, N}{St}\right)\, D^2-\frac{i\, M\, N'}{St}\, D -k\, \left(U''+k^2\, U-\frac{i\, k\, M\,N}{St} \right)$, $l_2 = N'+N\,D$ and $l_3 = i\, k\, U+\frac{1}{St}$. We solve the system of equations (\ref{eqn4p11}) numerically, for a simple shear flow ($U = y$) with a smooth base state number density profile as given by equation (\ref{eqn4p10}), with decaying boundary conditions for all eigenfunctions in the farfield.

\begin{figure}
\centerline{\includegraphics[width=\linewidth]{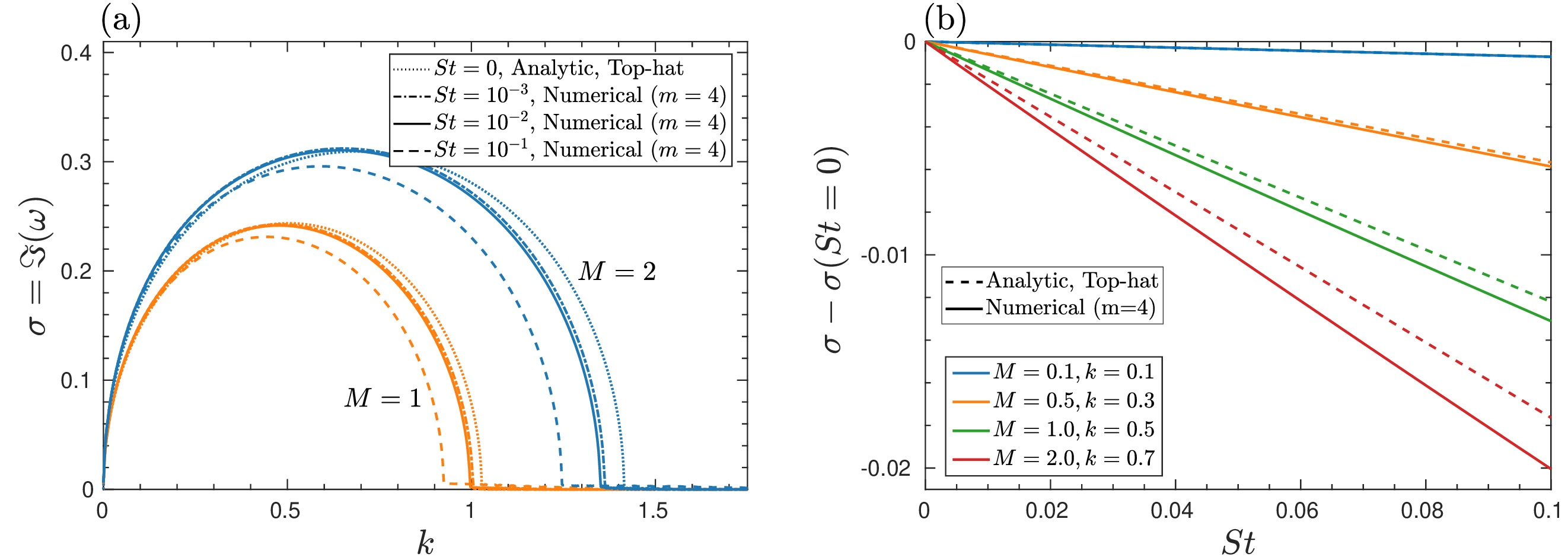}}
  \caption{(a) Growth rate $\sigma$ versus $k$ for two different mass loadings, $M=1$ (orange) and $M=2$ (blue), for various $St$ values obtained numerically for a smooth number density profile ($m=4$). The analytical result for the $St=0$ case with a top-hat number density profile is also shown for comparison. (b) The difference in growth rate for a finite $St$ case from the $St=0$ case, plotted versus $St$ for various combinations of $M$ and $k$. Here, $k$ is chosen to be approximately the optimum wavenumber to the respective $M$ value. Continuous lines represent the numerical results for a smooth profile ($m=4$), while dashed lines represent the analytic asymptotes for a top-hat profile.}
\label{fig8}
\end{figure}
\begin{figure}
\centerline{\includegraphics[width=\linewidth]{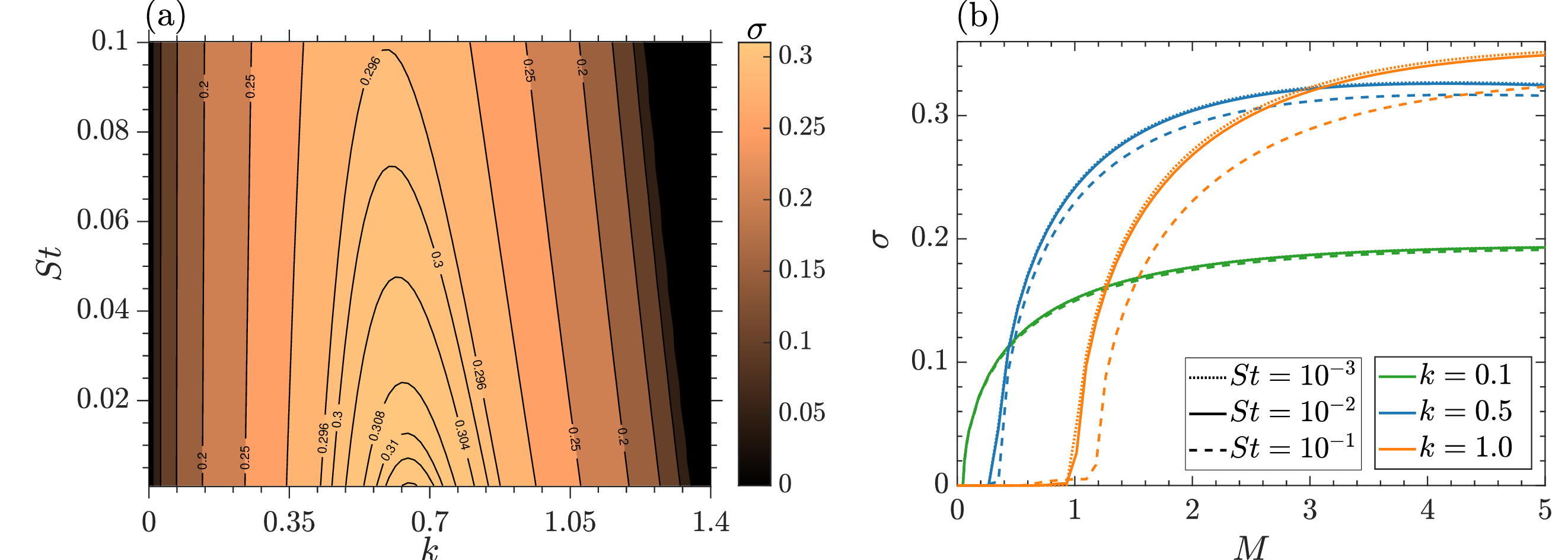}}
  \caption{(a) A contour plot of growth rate $\sigma$ in the $k$ versus $St$ plane for a mass loading of $M = 2$. (b) The variation of growth rate versus mass loading for various combinations of $k$ and $St$ values: dotted line represents $St = 10^{-3}$, continuous line represents $St = 10^{-2}$, and dashed line represents $St = 10^{-1}$. The results are obtained numerically for a smooth base state number density profile with $m=4$.}
\label{fig9}
\end{figure}

The results obtained are presented in figure \ref{fig8}. Figure \ref{fig8}(a) depicts the growth rate versus $k$ curve for various $St$ values, considering a smooth number density profile with $m=4$, for two different mass loading values. The corresponding curve for $St=0$ with a top-hat number density profile is also displayed for comparison. It is evident that as $St$ increases, the growth rate decreases for all mass loading values and wavenumbers. Furthermore, the cut-off wavenumber $k_\textrm{cutoff}$ decreases with increasing $St$. However, upon closer inspection, it is observed that with a finite Stokes number, the instability persists for $k > k_\textrm{cutoff}$, albeit with a significantly lower growth rate. In figure \ref{fig8}(b), the deviation of the growth rate corresponding to finite $St$ from the $St=0$ case is plotted as a function of Stokes number for various mass loading values. The wavenumbers are chosen to correspond to the fastest-growing modes for the corresponding mass loading values. The results are obtained numerically for a smooth number density profile of $m=4$. Analytically obtained asymptotes for the respective cases with top-hat number density profiles are plotted for comparison. The figure demonstrates that the growth rate decreases as the Stokes number increases.

Figure \ref{fig9}(a) presents a contour plot of the growth rate in the $k$ versus $St$ plane for a mass loading $M=2$, obtained numerically, for a smooth number density profile of $m=4$. The plot illustrates that the growth rate decreases with $St$ and increases with $M$. This implies that particle inertia stabilises the instability, whereas the mass loading (representing the two-way coupling strength) has a destabilizing effect. This qualitative trend would also hold for other mass loading values, though they are not shown here. In figure \ref{fig9}(b), the variation of growth rate with mass loading is depicted for various combinations of $k$ and $St$. This plot reiterates the deduction about the dampening of instability by $St$ and the strengthening of instability by $M$. It is important to note that the instability persists even in the limit of $St = 0$, whereas it vanishes in the $M=0$ limit (i.e., in the one-way coupling case). The analysis presented in this section confirms the existence of an instability in the system under consideration. The source of instability is attributed to the two-way coupling between the particles and the fluid, and it is observed that the instability can be dampened by the particle inertia. 
\section{Comparison of EL results with LSA results}
\label{sec5}
In the previous sections, we demonstrated the existence and mechanism of the instability using a linearized continuum model for the system. What remains is a quantitative comparison of the predictions from linear stability analysis with fully non-linear EL simulations.
We conducted a series of EL simulations across various parameter ranges to obtain the dispersion curve, maintaining most settings as described in \S \ref{sec2p2}. In these simulations, the Stokes number ($St$) was kept constant at $10^{-3}$. To ensure a fair comparison with the inviscid LSA, the EL simulations were carried out with a fixed $Re_{L_x} = 5000$, which is sufficiently high to make viscous effects negligible. The box domain size was set as $L_x = 2\, \pi \, \textrm{m}$, $L_y = 6\, \pi \, \textrm{m}$ and $L_z=3\, d_p$. The average volume fraction $\langle \phi_p \rangle$ of particles within the band was adjusted to achieve various mass loading values. To confine only one wave within the domain in the $x$ direction, we fixed the dimensional wavenumber to $k = 1\, \textrm{m}^{-1}$. By adjusting the bandwidth $h$, we obtained different non-dimensional $k$ values. In all selected cases, the ratio $h/d_p$ was set to be large (on the order of $10^2$ to $10^3$), ensuring that fluctuations caused by discrete particle forcing remained well below the viscous dissipation scale. This configuration enabled fluid-particle coupling primarily through collective particle dynamics at scales comparable to $h$, rather than discrete effects at inter-particle distances. Since $L_x$ is fixed, varying $h$ leads to a change in the ratio $L_x/h$ across the range from $2.5\, \pi$ to $32\, \pi$. However, $L_y$ remains sufficiently large in all these cases to avoid any possible interactions with other periodic simulation boxes in the $y$ direction.
\begin{figure}
\centerline{\includegraphics[width=1.2\linewidth]{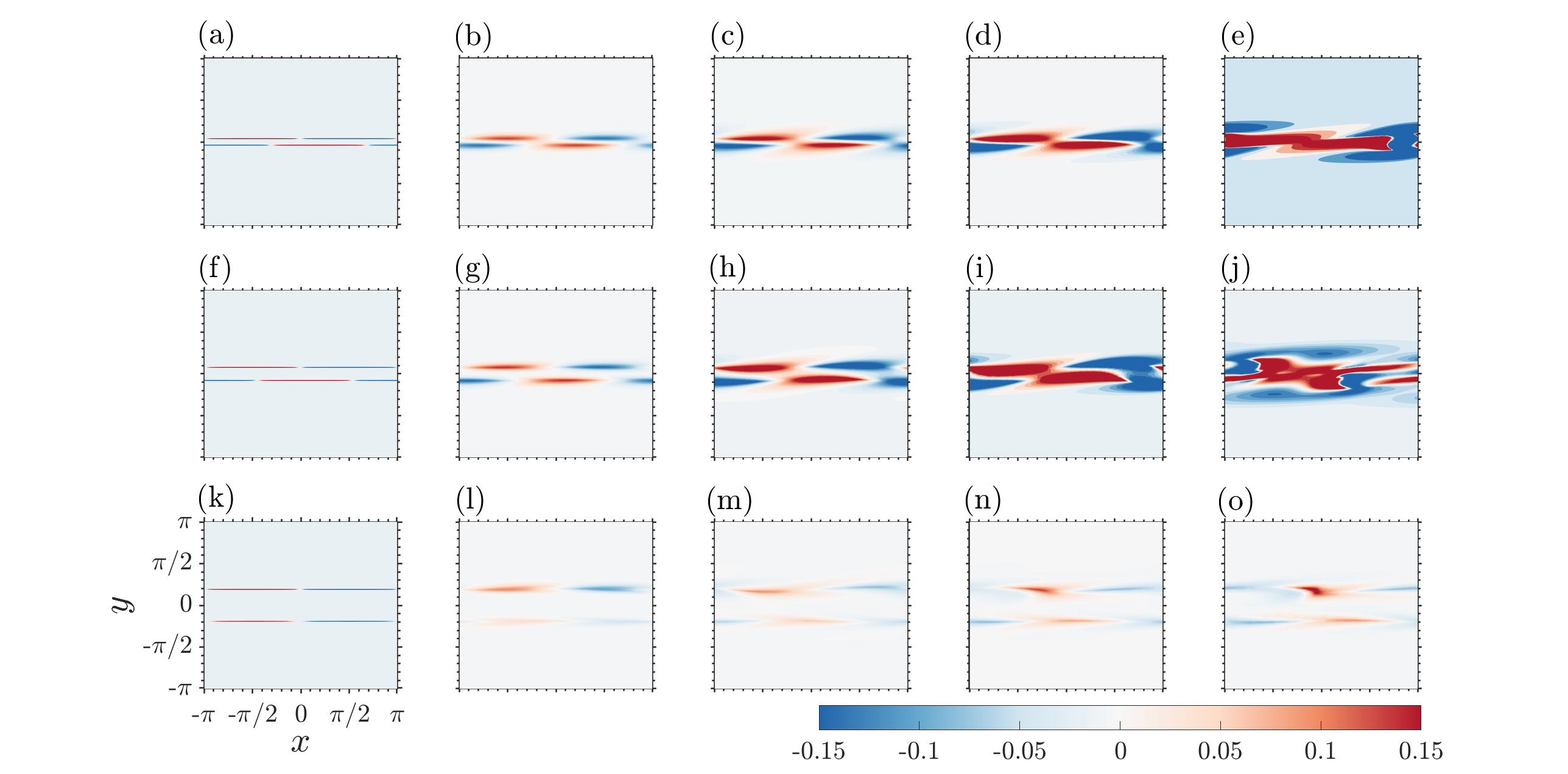}}
  \caption{The time evolution of the perturbation vorticity field $\tilde{q}_z$ (scaled by its initial maximum value) for a two-way coupled simulation with $M=1$ and $St=10^{-3}$ is plotted for three different non-dimensional wavenumbers: (a-e) top panel $k = 0.25$, (f-j) middle panel $k = 0.5$, and (k-o) bottom panel $k = 1.2$. The non-dimensional time stamps are as follows: (a,f,k) $t = 0$, (b,g,l) $t = 1$, (c,h,m) $t = 5$, (d,i,n) $t=9$, and (e,j,o) $t = 20$. The initial vorticity perturbation used is the eigenmode corresponding to a top-hat number density profile with $St = 0$, obtained analytically. The colour scale is constrained to $[-0.15, 0.15]$ for easy visualization. The spatial units in the plots are dimensional.}
\label{fig10}
\end{figure}
\begin{figure}
\centerline{\includegraphics[width=1.2\linewidth]{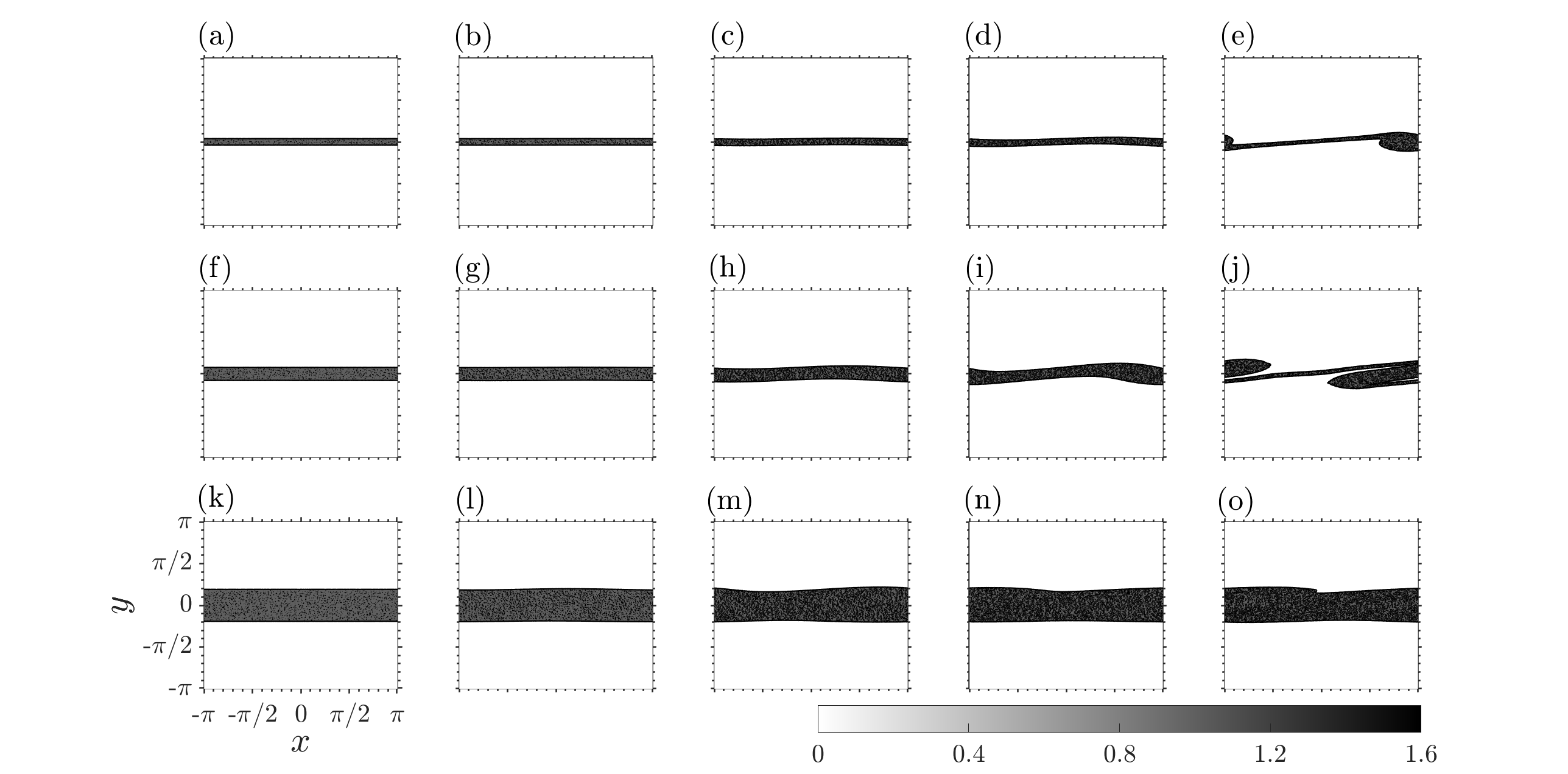}}
  \caption{The evolution of the total particle number density field $ \phi_p/\langle \phi_p\rangle$ for the same set of simulations in figure \ref{fig10}. }
\label{fig11}
\end{figure}

The initial conditions of the numerical simulations consist of a combination of the base state (a simple shear flow with a top-hat number density profile) and the perturbation eigenmodes obtained from the corresponding LSA. To perturb our system, we utilize the analytic expressions for the eigenmodes of the velocity field obtained for the corresponding $St = 0$ case. The initial particle number density field is left unperturbed since the eigenmodes are theoretically singular Dirac delta functions. Although this initial arrangement may not correspond precisely to an eigenmode of the system under consideration (due to the omission of finite $St$ aspects and perturbations in the number density field), it will closely resemble the eigenmode. As time progresses, the applied perturbations are expected to excite the most unstable mode, i.e., the respective eigenmode of the system. The initial perturbation fields used for the flow field can be visualized in the representative figure \ref{fig10}(a, f, k), and the corresponding initial particle distribution can be seen in figure \ref{fig11}(a, f, k). To ensure that the simulations capture the linear regime described by LSA and to delay any nonlinear effects, the perturbations are initialized with a small relative amplitude $\epsilon = 10^{-2}$. 
The location of the particles are randomly initialised within each cell inside the particle band, defined by $\lvert y \rvert \leq h/2$.

Figure \ref{fig10} illustrates the time evolution of the perturbation vorticity field (normalised by its initial maximum value) for a case with $M=1$ and $St = 10^{-3}$, depicted for three different non-dimensional wavenumbers: $k=0.25$ (top panel), $k=0.5$ (middle panel), and $k=1.2$ (bottom panel). As earlier, the plots are presented in a dimensional spatial domain of $2\pi \times 2\pi$ for better visualisation. The variation in wavenumbers is noticeable from the variation in $h$ in the initial disturbance fields (panels a,f,k). The initial evolution of the perturbation fields confirms the excitation of the respective eigenmodes. For the smaller wavenumber (top panel), it can be seen that the perturbations grow over time, and the vorticity field amplifies and spreads across a larger region, indicating the presence of instability. At $k=0.5$ (middle panel), the instability becomes more pronounced and transitions to the nonlinear regime rapidly, as expected theoretically, since the maximum growth corresponds to a non-dimensional wavenumber closer to $k=0.5$. The theory predicts no linear instability for the larger wavenumber (bottom panel), which is also evident from the snapshots where there is minimal growth of the vorticity field. Figure \ref{fig11} shows the corresponding evolution of the total particle number density field. It can be observed that although the number density field is initially not perturbed, eventually, the eigenmodes are getting excited. Additionally, it can be noticed that the evolution of the number density field is coherent with the corresponding perturbation vorticity field at all times for all wavenumbers. This correspondence also holds theoretically for a top-hat profile, as we have already seen using LSA that the vorticity perturbation and number density perturbation fields both are Dirac delta functions. 
The snapshots show that the case with wave number $k=0.5$ transitions to the nonlinear regime much faster than the others.

In order to determine the total perturbation growth rates from the present simulations, we calculate the disturbance kinetic energy associated with the perturbations as
\begin{equation}
    E = \int  \frac{1}{2}(\textbf{u}-\textbf{U})\cdot (\textbf{u}-\textbf{U})\, dV~,
    \label{eqn5p1}
\end{equation}
where $\textbf{U} = \Gamma\, y\, \hat{\textbf{x}}$ is the base state simple shear flow, is subtracted from $\textbf{u}$ - the total fluid velocity obtained from EL simulations, to obtain the disturbance fluid velocity. The integration is performed over the entire volumetric domain of the periodic simulation box.
\begin{figure}
\centerline{\includegraphics[width=1.1\linewidth]{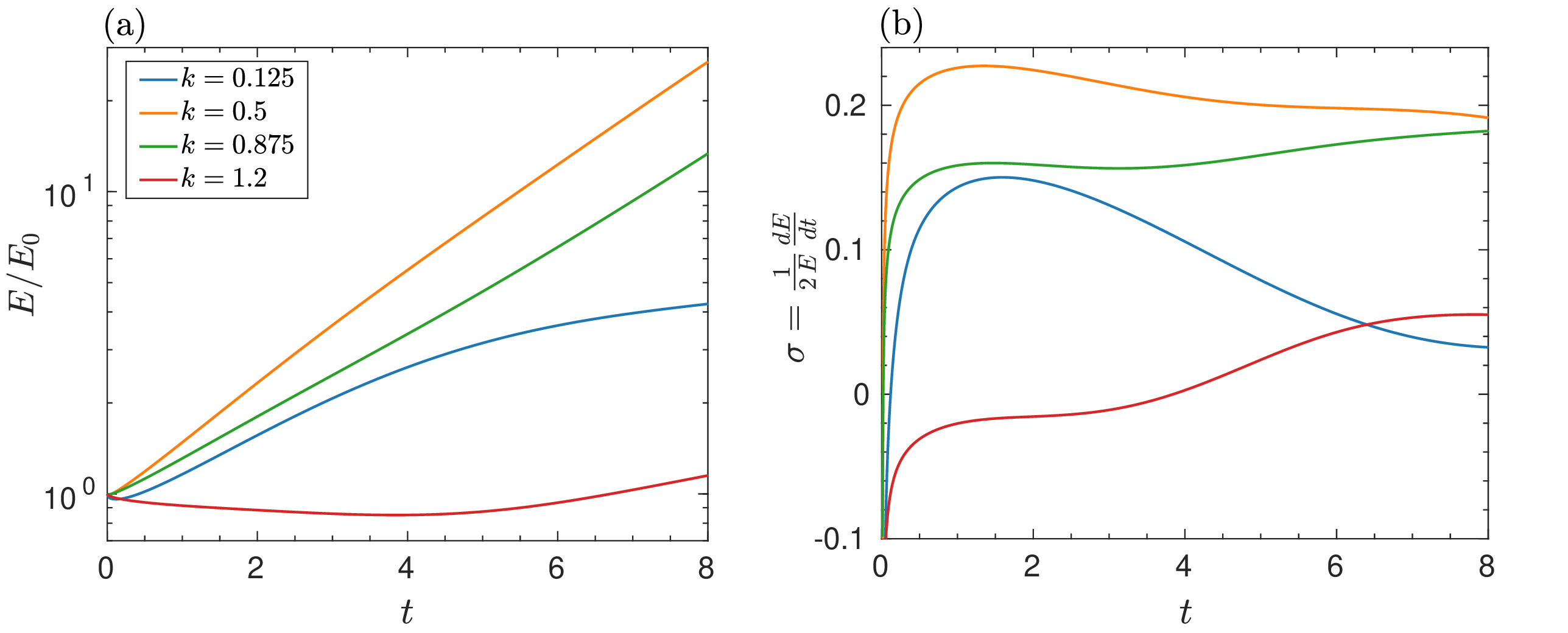}}
  \caption{The evolution of (a) total perturbation energy $E$ (normalized by its initial value) and (b) the estimate for the instantaneous growth rate with time, obtained from EL simulations, for 
  mass loading $M=1$,
   for various $k$ values.}
\label{fig12}
\end{figure}

\begin{figure}
\centerline{\includegraphics[width=1.1\linewidth]{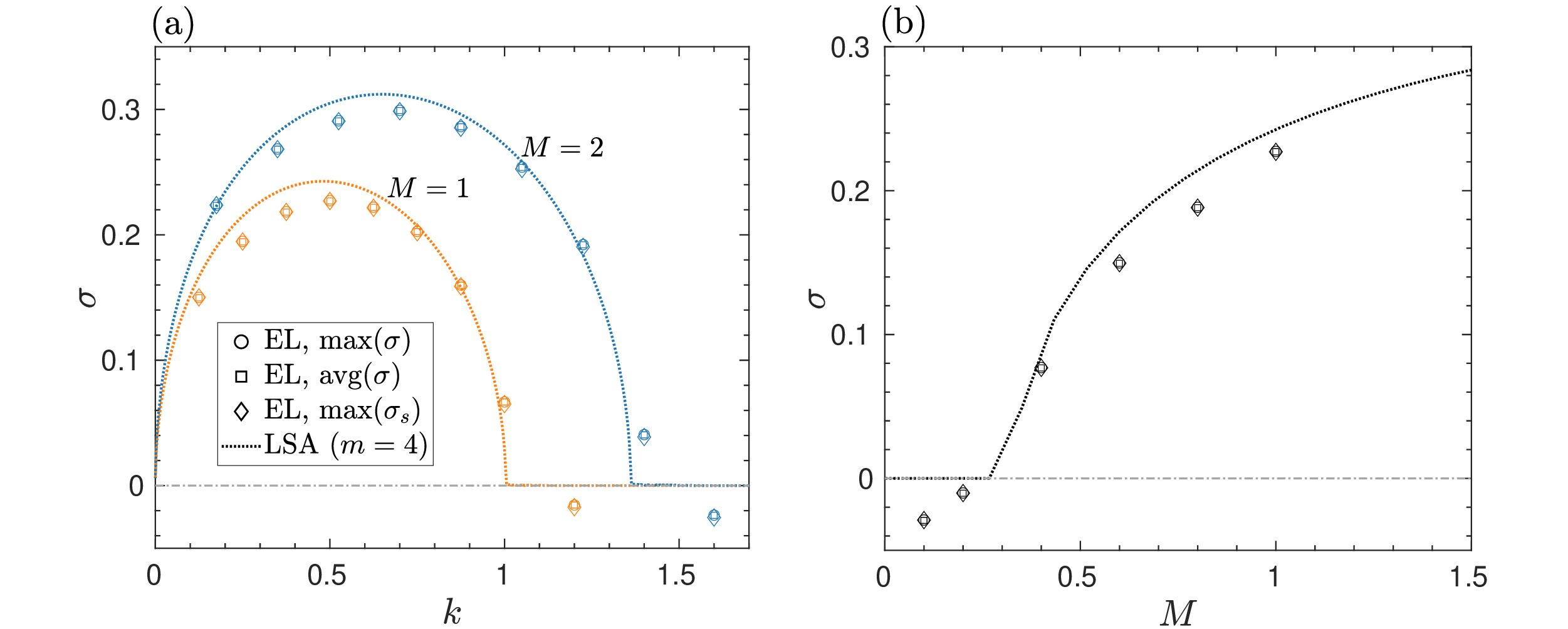}}
  \caption{The growth rate $\sigma$ obtained from EL simulations for particles with $St = 10^{-3}$ is plotted (a) against $k$ for two mass loadings $M=1$ (orange colour) and $M=2$ (blue colour), and (b) against $M$ for $k = 0.5$. Various estimates for the growth rate are indicated by different markers: $\circ$ denotes the peak value of the growth rate ($\textrm{max}(\sigma)$) estimated from the total perturbation energy, $\square$ represents the average value of the growth rate ($\textrm{avg}(\sigma)$) in a time window $\Delta t = 0.5$ about the peak growth rate of the total perturbation energy, and $\diamond$ indicates the peak growth rate obtained from the energy of the seeded mode ($\textrm{max}(\sigma_s)$). For comparison, the corresponding curves numerically obtained from LSA for a smooth number density profile of $m=4$ are also shown using dotted lines.}
\label{fig13}
\end{figure}

Figure \ref{fig12}(a) shows the evolution of total perturbation energy ($E$), scaled with the initial value ($E(t=0) = E_0$), computed from EL simulations, presented in a semi-logarithmic scale for various non-dimensional wavenumbers $k$, corresponding to 
a mass loading of $M=1$.
The plots exhibit a linear trend at initial times, indicating an exponential dependence on time. If the growth rate of velocity perturbations is denoted as $\sigma$, then the growth rate of energy perturbations would be $2\, \sigma$. Based on this argument, one may derive an expression to assess the growth rate $\sigma$ from EL simulations as
 \begin{equation}
     \sigma = \frac{1}{2E}\, \frac{d E}{d t}~.
     \label{eqn5p2}
\end{equation}
Figure \ref{fig12}(b) shows the time evolution of the growth rate estimated for EL simulation results using the expression given in equation (\ref{eqn5p2}). Unlike the prediction from LSA, which suggests a constant value, the instability growth rate exhibits variation over time in EL simulations. The figure indicates that the growth rate peaks between $t = 0$ and $t=2$, depending on the seeded mode. The initial transient behaviour may attributed to imperfections in the initial conditions.
The time window surrounding the peak growth rate in figure \ref{fig12}(b) corresponds to the exponential growth of kinetic energy observed in figure \ref{fig12}(a), thereby closely aligning with the dynamics predicted by LSA. Eventually, the growth rate decreases as the perturbation kinetic energy saturates, with the possibility of later increase due to nonlinear effects. 
An estimate for comparison with LSA would be the peak value of the growth rate for EL simulations obtained from equation (\ref{eqn5p2}). The non-dimensional growth rates $\sigma$ are thus obtained and plotted against the corresponding non-dimensional wavenumber $k$ for two different mass loading values in figure \ref{fig13}(a), represented by circle markers. The growth rate numerically obtained from LSA for a smooth number density profile ($m=4$) is also compared. The results show good agreement, although some discrepancies are observed in the numerical values. Considering that we are comparing completely different simulation classes: fully nonlinear EL simulations with LSA of the Euler-Euler model for the system under consideration, these differences are expected. Furthermore, the growth rate obtained from EL simulations indicates that energy perturbations decay beyond the critical wavenumber, contrary to what linear theory predicts. This could be attributed to the viscous effects in EL simulations.

In order to quantify the contribution of a specific wave mode $k$ to the overall dynamics, we compute the kinetic energy associated with this mode using Fourier decomposition as
\begin{equation}
    E_k =  \int_{-L_z/2}^{L_z/2}\int_{-L_y/2}^{L_y/2}\,  \frac{1}{2}\, \overline{\hat{\textbf{u}}_k(y,z,t)} \cdot \hat{\textbf{u}}_k(y,z,t)\, dy\, dz~,
    \label{eqn5p3}
\end{equation}
where $\overline{(\cdot)}$ denotes complex conjugation, and $\hat{\textbf{u}}_k$ is the Fourier amplitude of the perturbation velocity field, given by
\begin{equation}
    \hat{\textbf{u}}_k(y,z,t) = \frac{1}{\sqrt{2\, \pi}}\, \int_{-L_x/2}^{L_x/2}\,  (\textbf{u}-\textbf{U})\, \exp(-i\, k\, x)\, dx~.
    \label{eqn5p4}
\end{equation}
The normalization for the Fourier amplitude is chosen such that Plancherel's identity yields $\int E_k \, dk = E$. The peak growth rate corresponding to the seeded eigenmode can thus be extracted using equations (\ref{eqn5p3}) with (\ref{eqn5p2}), and are also shown in figure \ref{fig13} using diamond markers. These growth rates are also adequately consistent, qualitatively and quantitatively, with the LSA results. Additionally, the figure shows an average estimate of the growth rate over a non-dimensional time window $\Delta t = 0.5$ centred around the peak value of total perturbation energy, depicted using square markers, which also compares well with the LSA results. The differences between these various growth rate estimates can also be observed to be very small.

Figure \ref{fig13}(b) displays the growth rate estimates from EL simulations for the $k=0.5$ case plotted for various mass loading values. As the mass loading increases, the growth rate also increases, indicating that the strength of the two-way coupling determines the strength of the instability. The trend in growth rate aligns well with the LSA results (shown as a dotted curve). Below a critical mass loading, however, the EL simulations show damped modes instead of the neutral modes observed in LSA, likely due to the inherent viscous nature of EL simulations, as mentioned earlier.
\section{Conclusion}
\label{sec6}
In this study, we have investigated a novel type of instability arising from the interaction between a simple shear flow and a non-uniform distribution of particles. Individually, each component is stable; a simple shear flow remains stable against infinitesimal perturbations, and a localised band of particles is stable without any gravitational or buoyancy effects. However, when considering the system as a whole, the momentum feedback from the particles to the fluid phase renders the system unstable. We utilised EL simulations to show the existence of instability, treating the fluid phase as a continuum and the particles as discrete using Eulerian and Lagrangian descriptions, respectively. The two-way coupling between the fluid and particle phases is crucial for the observed instability. The strength of the momentum feedback from particles to the fluid is directly proportional to the particle mass fraction and inversely proportional to particle inertia. Consequently, increasing the particle mass fraction enhances the instability while increasing particle inertia suppresses it. The simulations were conducted at high Reynolds numbers, indicating that the instability originates from inviscid effects. Moreover, we conducted simulations where we disabled the momentum feedback from particles to the fluid, while retaining the momentum exchange from the fluid to the particle phase. As anticipated, the system remained stable under these conditions, with particles solely advected by the shear flow.

Furthermore, a continuum model of the particle phase is employed to illustrate this instability's modal nature and explain its generation mechanism. In the semi-dilute limit, the particle phase can be modelled as a separate pressureless fluid. The model adopts an Eulerian approach for particle and fluid phases, known as the two-fluid model. Linear stability analysis (LSA) of this model was conducted under the $St = 0$ and $St \ll 1$ limits to quantify the growth rate of instability, utilising analytical and numerical techniques. The analytical investigation primarily focused on a simple top-hat distribution of particles, while a numerical treatment was reserved for smooth, localised particle distributions. In both scenarios, the analysis demonstrated that instability emerges as the momentum feedback from particles to the fluid phase is considered. Surprisingly, the instability persists even in the $St \rightarrow 0$ limit, suggesting that although it originates from particle inertia, its existence does not solely depend on inertia when the fluid-particle coupling is accounted for. However, increasing $St$ results in a damping effect on the growth rate of instability. The growth rate estimates from EL simulations matched the LSA results reasonably well.

To gain a mechanistic view of the instability, we examine the regime of small particle inertia ($St \rightarrow 0$). In this limit, the two-fluid model simplifies into a single-fluid model that describes the continuum evolution of a fluid with an effective density due to the suspended particles. The non-uniformity in particle distribution is thus reflected in the effective density of this fluid. The system now resembles a density-stratified fluid capable of supporting propagating edge waves at interfaces where vorticity or density exhibits discontinuities. Our system has two density interfaces, which individually support propagating edge waves generated due to non-Boussinesq baroclinic effects in isolation in a background shear flow. Unlike in the case of classical edge waves, the baroclinic torque arises from the misalignment between base state density stratification and disturbances in fluid acceleration. The propagating waves at both interfaces feel each other's presence once brought closer, and they interact through phase locking and mutual amplification, culminating in the instability observed.

The present analysis has made several assumptions, and the relaxation of many of those would augment the instability found in the present study. The background flow is a simple shear flow, thus disallowing any shear instabilities that might arise \citep{drazin2004hydrodynamic}. We have also neglected the effects of gravitational settling, non-zero fluid inertia, concentration-dependent viscosity and hydrodynamic interactions. As was discussed earlier, in a simple shear flow of uniformly distributed negatively buoyant inertial particles, velocity perturbations can lead to preferential concentration of particles in regions of low vorticity. \cite{kasbaoui2015preferential} showed that the variation of gravitational force exerted on the suspension thus induced would result in an algebraic instability. Ignoring gravitational effects, if the effects of fluid inertia were  included instead, the particles would experience lift forces. \cite{drew1975lift} showed that the lift force can lead to an instability in a bounded simple shear flow. A departure from the diluteness approximation would manifest first as a concentration-dependent viscosity and then as non-zero particle pressure. With the former physics, non-uniformity in particle concentration can lead to viscosity contrasts in the suspension, potentially causing short-wave viscous instabilities \citep[see, e.g.,][]{hooper1983shear,hinch1984note,sahu2011linear}. The latter, particle pressure due to hydrodynamic interactions, is responsible for shear-induced migration in dense suspensions \citep{leighton1987shear}. Shear-induced migration would be absent in a simple shear flow. However, the perturbation-induced migration could lead to a short-wave instability depending on the importance of concentration-driven diffusion relative to shear-induced migration \citep{goddard1996migrational}. Coupling even some of the physics mentioned above with the novel instability found in the present study would pave the way for a better understanding of instabilities in dusty shear flows and how they can have fundamentally different transition routes than their particle-free counterparts.





\backsection[Funding]{A.R. acknowledge SERB project CRG/2023/008504 for funding. A.V.S.N. thanks the Prime Minister’s Research Fellows (PMRF) scheme, Ministry of Education, Government of India. A.R. and A.V.S.N. acknowledge the support of the Geophysical Flows Lab (GFL) at IIT Madras. M.H.K acknowledges support from the US National Science Foundation (award \#2148710, CBET-PMP).}


\backsection[Declaration of interests]{The authors report no conflict of interest.}


\backsection[Author ORCIDs]{A. V. S. Nath, https://orcid.org/0000-0003-2144-2978; A. Roy, https://orcid.org/0000-0002-0049-2653; M. H. Kasbaoui, https://orcid.org/0000-0002-3102-0624.}


\appendix
\section{Linearisation and normal mode analysis}
\label{appA}
The linearised perturbation equations (\ref{eqn4p2}) can be expanded in component form as,
\begin{subeqnarray}
\frac{\partial \tilde{u}_x}{\partial x}+\frac{\partial \tilde{u}_y}{\partial y} = 0~,\\
   \frac{\partial \tilde{n}}{\partial t}+N\, \left(\frac{\partial \tilde{v}_x}{\partial x}+\frac{\partial \tilde{v}_y}{\partial y} \right)+\tilde{v}_y\, N'+U\, \frac{\partial \tilde{n}}{\partial x} = 0~,\\
    \frac{\partial \tilde{u}_x}{\partial t}+U\, \frac{\partial \tilde{u}_x}{\partial x}+\tilde{u}_y\, U' = -\frac{\partial \tilde{p}  }{\partial x}+\frac{M\, N}{St}\, (\tilde{v}_x-\tilde{u}_x)~,
    \\
    \frac{\partial \tilde{u}_y}{\partial t}+U\, \frac{\partial \tilde{u}_y}{\partial x}=-\frac{\partial \tilde{p}}{\partial y}+\frac{M\, N}{St}\, (\tilde{v}_y-\tilde{u}_y)~,\\
\frac{\partial \tilde{v}_x}{\partial t}+U\, \frac{\partial \tilde{v}_x}{\partial x}+\tilde{v}_y\, U' = \frac{\tilde{u}_x-\tilde{v}_x}{St}~,\\
    \frac{\partial \tilde{v}_y}{\partial t}+U\, \frac{\partial \tilde{v}_y}{\partial x}=\frac{\tilde{u}_y-\tilde{v}_y}{St}~.
    \label{eqnA1}
\end{subeqnarray}
To analyze stability, we need to solve the system of equations (\ref{eqnA1}) for the perturbation quantities $\{\tilde{u}_x,\tilde{u}_y,\tilde{p},\tilde{v}_x,\tilde{v}_y,\tilde{n}\}$. Assuming a modal form for perturbations $\{\tilde{u}_x,\tilde{u}_y,\tilde{p},\tilde{v}_x,\tilde{v}_y,\tilde{n}\} = \{\hat{u}_x,\hat{u}_y,\hat{p},\hat{v}_x,\hat{v}_y,\hat{n}\}(y)\, \exp\{i\, (k\, x-\omega\, t)\}$, we obtain the linearized equations in modal space as
\begin{subeqnarray}
\hat{u}_x = \frac{i}{k}\, \frac{d \hat{u}_y }{d y}~,\\
    i\, k\, (U-c)\, \hat{n}+N\, \left(\frac{d \hat{v}_y}{d y}+i\, k\, \hat{v}_x \right)+\hat{v}_y\, N' = 0~,\\
    i\, k\, \hat{u}_x\, (U-c)+U'\, \hat{u}_y=-i\, k\, \hat{p}+\frac{M\, N}{St}\, (\hat{v}_x-\hat{u}_x)~,
    \\
    i\, k\, \hat{u}_y\, (U-c)=-\frac{d \hat{p}}{d y}+\frac{M\, N}{St}\, (\hat{v}_y-\hat{u}_y)~,\\
i\, k\, \hat{v}_x\, (U-c)+U'\, \hat{v}_y=\frac{\hat{u}_x-\hat{v}_x}{St}~,\\
    i\, k\, \hat{v}_y\, (U-c)=\frac{\hat{u}_y-\hat{v}_y}{St}~.
    \label{eqnA2}
\end{subeqnarray}
After eliminating $\hat{u}_x$ and $\hat{p}$ from equation (\ref{eqnA2}c) using (\ref{eqnA2}a) and (\ref{eqnA2}d), we obtain a single equation for the evolution of $\hat{u}_y$ as
\begin{eqnarray}
    \left(U-c-\frac{i\, M\, N}{k\, St}\right)\, \frac{d^2 \hat{u}_y }{d y^2}-\frac{i\, M\, N'}{k\, St}\, \frac{d \hat{u}_y}{d y}- \left(  U''+k^2\, (U-c)-\frac{i\, k\, M\, N}{St}\right)\, \hat{u}_y \nonumber \\
    +\frac{M\, N}{St}\, \frac{d \hat{v}_x}{d y} +\frac{M\, N'}{St}\, \hat{v}_x -\frac{i\, k\, M\, N}{St}\, \hat{v}_y = 0~.
    \label{eqnA3}
\end{eqnarray}
Equations (\ref{eqnA2}e-f) can be solved simultaneously to obtain $\hat{v}_x$ and $\hat{v}_y$ in terms of $\hat{u}_x$ and $\hat{u}_y$ as $\hat{v}_x = \hat{u}_x\, \Lambda-St\, U'\, \hat{u}_y \, \Lambda^2$ and $\hat{v}_y = \hat{u}_y\, \Lambda$, where $\Lambda(y) = \left(1+i\, k\, St\, (U(y)-c)\right)^{-1}$. Substituting them into equations (\ref{eqnA3}) and (\ref{eqnA2}b) gives the nonlinear eigenvalue system 
\begin{subeqnarray}
    \left\{ (U-c)\, (1+M \, N\,\Lambda) \right\} \,D^2\hat{u}_y+\left\{(U-c)\, M\, N'\, \Lambda\right\} \, D\hat{u}_y \nonumber \\
    -\left\{U''+k^2\,(U-c)\, (1+M \, N\,\Lambda) +M\, D\left[ N\, \Lambda^2\, U' \right] \right\}\, \hat{u}_y = 0~,\\
    i\, k\, (U-c)\, \hat{n}+\left[ \Lambda\, N' -2\, i\, k\, St\, \Lambda^2\, N\, U'\right] \hat{u}_y=0~.
    \label{eqnA4}
\end{subeqnarray}
By defining $\mathcal{U}= (U-c)\,(1+ M\, N\, \Lambda)$, the system can be further simplified as in equation (\ref{eqn4p3}). Integrating equation (\ref{eqn4p3}a) across a discontinuous interface $y = a$ yields the interface boundary condition as
\begin{equation}
 \lBrack\hat{u}_y\, \mathcal{U}'-\mathcal{U}\,  D\hat{u}_y-(U-c) \, M\,N'\, \Lambda \,\hat{u}_y\rBrack_{y = a^-}^{y = a^+} + k^2\, \cancelto{0}{\int_{a^-}^{a^+}\mathcal{U}\, \hat{u}_y\, dy}= 0~,
 \label{eqnA5}
\end{equation}
where the last integral vanishes, assuming the discontinuity in $U$ or $N$ or both can never be more than a finite jump. Here $\lBrack \cdot \rBrack_{y = a^-}^{y = a^+}$ denotes the difference of the quantity inside the square bracket evaluated at infinitesimally above and below the interface $y=a$.  
\section{The dusty Rayleigh criterion considering non-Boussinesq baroclinic effects}
\label{appB}
Consider the equation (\ref{eqn4p5}), which incorporates non-Boussinesq baroclinic effects but neglects gravitational baroclinic effects. Divide throughout by $(U-c)$, assuming $U \neq c$. After rearranging and combining terms, we obtain
\begin{equation}
    D[\rho\, D\hat{u}_y]-\frac{D[\rho\, U']}{(U-c)}\, \hat{u}_y-k^2\, \rho\, \hat{u}_y = 0~.
    \label{eqnB1}
\end{equation}
After multiplying throughout by the complex conjugate $\overline{\hat{u}_y}$ and integrating over the entire $y$ domain $\mathscr{D}$, we get
\begin{equation}
    \int_{\mathscr{D}} \overline{\hat{u}_y}\, D[\rho\, D\hat{u}_y]\, dy-\int_{\mathscr{D}} \left(\frac{D[\rho\, U']}{(U-c)}\, +k^2\, \rho\ \right)\, \lvert \hat{u}_y \rvert^2\, dy = 0~.
\end{equation}
Using integration by parts in the first integral, we have
\begin{equation}
    \left[ \overline{\hat{u}_y}\, \rho\, D\hat{u}_y\right] -\int_{\mathscr{D}}  \rho\, \lvert D\hat{u}_y\rvert^2\, dy-\int_{\mathscr{D}} \left(\frac{D[\rho\, U']}{(U-c)}\, +k^2\, \rho\ \right)\, \lvert \hat{u}_y \rvert^2\, dy = 0~,
    \label{eqnB3}
\end{equation}
where the first term in the square bracket needs to be evaluated at the boundaries, and their difference needs to be taken. Since the perturbations should vanish at the boundaries (whether the domain is bounded or unbounded), this term evaluates to zero. After expanding $ c = c_R + i\, c_I $, the imaginary part of the remaining integral terms yields
\begin{equation}
i\, c_I\,\int_{\mathscr{D}} \frac{D[\rho\, U']}{\lvert U-c\rvert^2}\, \, \lvert \hat{u}_y \rvert^2\, dy = 0~.
\label{eqnB4}
\end{equation}
For the integral to be satisfied, the only possibility is that the term $ D[\rho\, U'] $ should have a sign change in the domain - a modified form of the Rayleigh criterion for instability. Note that this will only be a necessary criterion, but not a sufficient one.  

When considering the gravitational baroclinic effects, using the same procedure, we obtain that there should be a sign change for the quantity
\begin{equation}
    D[\rho\, U']+\frac{2\, g\, (U-c_R)\, \rho'}{(U-c_R)^2+c_I^2}~,
\end{equation}
in the domain. Here $g$ represents the acceleration due to gravity (appropriately scaled).

The real part of the equation (\ref{eqnB3}) yields,
\begin{equation}
     \int_{\mathscr{D}} \frac{D[\rho\, U']}{\lvert U-c\rvert^2}\,  \, (U-c_R)\lvert \hat{u}_y \rvert^2\, dy = -\int_{\mathscr{D}}  \rho\,\left\{  \lvert D\hat{u}_y\rvert^2 +k^2\, \lvert \hat{u}_y \rvert^2 \right\}\, dy~.
\end{equation}
Since the right-hand side of the equation is clearly negative, the left-hand side integral should also be negative, implying that among the integrands, $(U-c_R) \, D[\rho\, U']$ should be negative somewhere within the domain. However, upon expanding the integrand, we already know that the integral term containing $c_R$ is zero when there is instability, as shown by equation (\ref{eqnB4}). Thus, $c_R$ can be replaced by any arbitrary constant value without affecting the result. However, the appropriate choice would yield a modified version of the Fjørtoft criterion for instability. Here, it would be $U^* = U(y^*)$, where $y^*$ is the location in the flow domain where $D[\rho\, U']$ changes sign. Then, the modified version of the Fjørtoft criterion states that, for instability, the necessary (but not sufficient) criterion is $(U-U^*) \, (\rho\, U')' < 0$ somewhere within the flow domain.
\section{LSA for $St=0$ particles inside a top-hat band}
\label{appC}
For a simple shear background flow with a top-hat number density profile, the governing equation (\ref{eqn4p5}) simplifies to
\begin{subeqnarray}
    (y-c)\,(D^2-k^2)\hat{u}_y = 0~,\\
    (y-c)\, \hat{n} = 0~.
    \label{eqnC1}
\end{subeqnarray}
For discrete modes, i.e., when $ y \neq c $, the solutions of equation (\ref{eqnC1}a) take on exponential form, as given in equation (\ref{eqn4p6}). The decaying boundary condition for the eigenfunctions $ \hat{u}_y $ at the far field is already incorporated in this solution. We utilize the interface conditions to determine the unknown amplitudes $ A_1 $ to $ A_4 $ and the eigenvalue $ c $. Generally, the rate of change in the interface displacement perturbation $ \tilde{\eta} $ is responsible for the vertical fluid velocity disturbance, i.e., $ \tilde{u}_y = \mathcal{D} \tilde{\eta}/\mathcal{D} t = \partial \tilde{\eta}/\partial t+U\, \partial \tilde{\eta}/\partial x $. In modal space, this implies $ \hat{u}_y = i\, k\, (U-c)\, \hat{\eta} $. Here, since the background shear flow is continuous across the interface, this relation demands the continuity of the vertical velocity perturbation as
  \begin{subeqnarray}
      \lBrack \hat{u}_y \rBrack_{y=1/2^-}^{y=1/2^+}=0~,\\
      \lBrack \hat{u}_y\rBrack_{y=-1/2^-}^{y=-1/2^+}=0~.
      \label{eqnC2}
  \end{subeqnarray}
  By integrating the equation (\ref{eqnB1}) multiplied with $(U-c)$, across each interface, we get the continuity criteria for pressure as well as
  \begin{subeqnarray}
      \lBrack \rho\left\{\hat{u}_y\, U'-(U-c)\, D\hat{u}_y \right\}\rBrack_{y=1/2^-}^{y=1/2^+} =0~,\\
      \lBrack \rho\left\{\hat{u}_y\, U'-(U-c)\, D\hat{u}_y \right\}\rBrack_{y=-1/2^-}^{y=-1/2^+} =0~.
      \label{eqnC3}
  \end{subeqnarray}
 After incorporating the solution form (\ref{eqn4p6}) into these criteria (\ref{eqnC2}) and (\ref{eqnC3}), we obtain a set of four algebraic equations, as given in matrix form (\ref{eqn4p8}). For a nontrivial solution, the determinant of the coefficient matrix should be zero, which gives the dispersion relation in (\ref{eqn4p9}). Since the system (\ref{eqnC1}a) is a homogeneous differential equation, the unknown constants $A_1$ to $A_4$ cannot be uniquely determined. By fixing one of the amplitudes (let us say $A_2$), all others can be obtained by solving any three out of the four equations in (\ref{eqn4p8}). For instance, we obtain
\begin{subeqnarray}
    \frac{A_1}{A_2}=\frac{e^k\,(\left(c+1/2\right) k-At)+At e^{-k} \left(\left(c+1/2\right)
   k+1\right)}{(At+1) \left(c+1/2\right)
   k}~,\\
   \frac{A_3}{A_2} = \frac{\left(c+1/2\right) k-At}{(At+1) \left(c+1/2\right)
   k}~,\\
   \frac{A_4}{A_2} =\frac{At e^{-k} \left(\left(c+1/2\right)
   k+1\right)}{(At+1) \left(c+1/2\right)
   k}~.
\end{subeqnarray}
The value for $c$ should be used from the dispersion relation (\ref{eqn4p9}). 

The trivial solution to equation (\ref{eqnC1}b) is $\hat{n} = 0$ for any discrete modes, everywhere except at the interfaces $y = \pm 1/2$. Thus, the general form for the solution should be a combination of Dirac delta functions about the interfaces. Using conservation of the total number density, one can obtain $\hat{n} \propto \delta(y-1/2)-\delta(y+1/2)$.
\section{LSA for $St \ll 1$ particles inside a top-hat band}
\label{appD}
In the limit of small Stokes numbers ($St \ll 1$), the governing equations (\ref{eqn4p3}) for a simple shear flow with a top-hat distribution of particles simplify to
\begin{subeqnarray}
\left. \begin{array}{cc}(D^2-k^2)\hat{u}_y = 0~, \\[0.75pt] \\
(y-c)\, \hat{n} = 0~,\end{array}\right\} \quad \textrm{for} \quad  \lvert y \rvert > 1/2
     \\
\left. \begin{array}{cc}\left(D^2-k^2 + \frac{2\, M\, \alpha}{(1+M)\, (y-c)}\right)\, \hat{u}_y + \textit{O}(\alpha^2) = 0~, \\[0.75pt]\\i\, k\, (y-c)\, \hat{n}-2\, \alpha\, \hat{u}_y + \textit{O}(\alpha^2) = 0~,\end{array}\right\} \quad \textrm{for} \quad  \lvert y \rvert < 1/2  
\label{eqnD1}
\end{subeqnarray}
where $\alpha = i\, k\, St$. The solution for eigen functions $\hat{u}_y$ can be obtained in each layer as,
 \begin{equation}
      \hat{u}_y = \begin{cases} 
      A_1\, e^{-k\, y}~, & y > 1/2 \\
      A_3\, \textrm{W}_{l,1/2}(2\, k\, (y-c))+A_4\, \textrm{M}_{l,1/2}(2\, k\, (y-c))~, & \lvert y  \rvert <  1/2 \\
      A_2\, e^{k\, y}~, & y < -1/2
   \end{cases}
   \label{eqnD2}
  \end{equation}
where $\textrm{W}$ and $\textrm{M}$ represents Whittaker functions and $l = M \, \alpha\, /(k\, (1+M))$. Note that the decaying boundary conditions for $\hat{u}_y$ are already enforced for the far field. Considering the system accurate up to $\textit{O}(\alpha)$, we omit terms of $\textit{O}(\alpha^2)$ in subsequent expressions. The continuity of $\hat{u}_y$ at the interfaces yields two equations as 
  \begin{subeqnarray}
A_1\, e^{-k/2} = A_3\, \textrm{W}_{l,1/2}(-2\, \omega^-)+A_4\, \textrm{M}_{l,1/2}(-2\, \omega^-)~,\\
A_2\, e^{-k/2} = A_3\, \textrm{W}_{l,1/2}(-2\, \omega^+)+A_4\, \textrm{M}_{l,1/2}(-2\, \omega^+)~,
\label{eqnD3}
  \end{subeqnarray}
  where $\omega^{\pm} = k\, (c \pm 1/2)$. Expanding the interface condition equation (\ref{eqnA4}) for the small $St$ limit ($\alpha \ll 1$), we obtain:
    \begin{subeqnarray}
      \left[ (1+M)\left\{\hat{u}_y-(y-c)\, D\hat{u}_y \right\} -M \, \alpha\, (y-c)\, \left\{2\, \hat{u}_y-(y-c)\, D\hat{u}_y \right\} \right]_{y=1/2^-} \nonumber \\
      = \left[ \left\{\hat{u}_y-(y-c)\, D\hat{u}_y \right\}\right]_{y=1/2^+}~, \\
            \left[ (1+M)\left\{\hat{u}_y-(y-c)\, D\hat{u}_y \right\} -M \, \alpha\, (y-c)\, \left\{2\, \hat{u}_y-(y-c)\, D\hat{u}_y \right\} \right]_{y=-1/2^+} \nonumber \\
      = \left[ \left\{\hat{u}_y-(y-c)\, D\hat{u}_y \right\}\right]_{y=-1/2^-}~,
      \label{eqnD4}
  \end{subeqnarray}
where the quantities inside the square brackets, indicated as $\left[ \cdot \right]_{y=a}$, needs to be evaluated at the appropriate interface locations $y=a$. Substituting the solution (\ref{eqnD2}) into the above conditions (\ref{eqnD4}) yields 
\begin{subeqnarray}
    A_1\, e^{-k/2}\, (1-\omega^-)+A_3\, \left\{ \psi^-\, \omega^-\, \textrm{W}_{l,1/2}(-2\, \omega^-)-\chi^-\, \textrm{W}_{l,3/2}(-2\, \omega^-)\right\} \nonumber \\
    +A_4\, \left\{\psi^-\, \omega^-\, \textrm{M}_{l,1/2}(-2\, \omega^-)+\frac{(1+l)\, \chi^-}{6}\, \textrm{M}_{l,3/2}(-2\, \omega^-) \right\}=0~,\\
        A_2\, e^{-k/2}\, (1+\omega^+)+A_3\, \left\{ \psi^+\, \omega^+\, \textrm{W}_{l,1/2}(-2\, \omega^+)-\chi^+\, \textrm{W}_{l,3/2}(-2\, \omega^+)\right\} \nonumber \\
    +A_4\, \left\{\psi^+\, \omega^+\, \textrm{M}_{l,1/2}(-2\, \omega^+)+\frac{(1+l)\, \chi^+}{6}\, \textrm{M}_{l,3/2}(-2\, \omega^+) \right\}=0~,
    \label{eqnD5}
\end{subeqnarray}
where $\psi^{\pm} = l^2\, (M+1)\, \omega^{\pm}$ and $\chi^{\pm} = (l-1)\, (M+1)\, \omega^{\pm}\, (1+l\, \omega^{\pm})$. For a nontrivial solution to the system of equations (\ref{eqnD3}) and (\ref{eqnD5}), we obtain the dispersion relation as
\begin{align}
\left[(l+1)\, \chi^+ \,\textrm{M}_{l,\frac{3}{2}}(-2\, \omega^+)+6\, s^+ \textrm{M}_{l,\frac{1}{2}}(-2 \,\omega^+)\right] \left[s^- \,\textrm{W}_{l,\frac{1}{2}}(-2\, \omega^-)-\chi^- \,\textrm{W}_{l,\frac{3}{2}}(-2 \,\omega^-)\right] -\nonumber \\
 \left[(l+1)\, \chi^- \,\textrm{M}_{l,\frac{3}{2}}(-2\, \omega^-)+6 \,s^- \,\textrm{M}_{l,\frac{1}{2}}(-2\, \omega^-)\right]    \left[s^+\, \textrm{W}_{l,\frac{1}{2}}(-2\, \omega^+)-\chi^+\, \textrm{W}_{l,\frac{3}{2}}(-2 \,\omega^+)\right] = 0~,
    \label{eqnD6}
 \end{align}
where $s^{\pm} = 1\pm \omega^{\pm}+\psi^{\pm}\, \omega^{\pm}$.
The constants can also be obtained as
\begin{subeqnarray}
    \frac{A_1}{A_2} &=& \textrm{T}^{-1}\,\chi^-  \,\left((l+1)\, \textrm{M}_{l,\frac{3}{2}}(-2 \,\omega^-)\, \textrm{W}_{l,\frac{1}{2}}(-2 \,\omega^-)+6 \,\textrm{M}_{l,\frac{1}{2}}(-2\, \omega^-)
   \, \textrm{W}_{l,\frac{3}{2}}(-2\, \omega^-)\right)~,\\
   \frac{A_3}{A_2}&=&\textrm{T}^{-1}\,e^{-k/2}  \left((l+1)\, \chi^-\, \textrm{M}_{l,\frac{3}{2}}(-2\, \omega^-)+6 \,s^- \,\textrm{M}_{l,\frac{1}{2}}(-2\,\omega^-)\right)~,\\
   \frac{A_4}{A_2}&=&\textrm{T}^{-1}\,6\, e^{-k/2} \,\left(\chi^-\, \textrm{W}_{l,\frac{3}{2}}(-2\, \omega^-)-s^- \,\textrm{W}_{l,\frac{1}{2}}(-2\, \omega^-)\right)~,
      \label{eqnD7}
\end{subeqnarray}
where
\begin{align}
    \textrm{T} = (\omega^++1)^{-1}\,\left[\left(6\, \psi^+ \,\omega^+ \,\textrm{M}_{l,\frac{1}{2}}(-2 \,\omega^+)+(l+1)\, \chi^+\, \textrm{M}_{l,\frac{3}{2}}(-2 \,\omega^+)\right)\, \left(s^-
  \, \textrm{W}_{l,\frac{1}{2}}(-2\, \omega^-) \right. \right.   \nonumber \\
   \left. \left.  -\chi^-\, \textrm{W}_{l,\frac{3}{2}}(-2\, \omega^-)\right)  
   -\left(\psi^+\, \omega^+\, \textrm{W}_{l,\frac{1}{2}}(-2\, \omega^+)-\chi^+\, \textrm{W}_{l,\frac{3}{2}}(-2\, \omega^+)\right) \right.   \nonumber \\
   \left. \left((l+1)\, \chi^- \textrm{M}_{l,\frac{3}{2}}(-2\, \omega^-)+6 \,s^- \,\textrm{M}_{l,\frac{1}{2}}(-2
   \,\omega^-)\right)\right]~.
   \label{eqnD8}
\end{align}
Note that these expressions (\ref{eqnD6})-(\ref{eqnD8}) are not properly obtained perturbation solutions. We made an $O(St)$ assumption in equation (\ref{eqnD1}b), but from equation (\ref{eqnD2}) onwards, the expressions are nonlinear in $St$ (or $l$), and we left track of the fact that the expressions are accurate only upto $\textit{O}(St)$. Thus, we may expand the dispersion relation (\ref{eqnD6}) as a series in $l$ along with the ansatz $\omega = \omega_0+l\, \omega_1+\textit{O}(l^2)$, and the resulting dispersion relation can be easily solved in each order. The leading order will give the solution for $\omega_0$, equivalent to the dispersion relation for the $St = 0$ case in equation (\ref{eqn4p9}) with $\omega_0 = c\, k$. Solving the next order will give the correction term $\omega_1$ as
\begin{align}
   \omega_1 =  \left[2 \omega_0 \left(e^k (M+2)^2-e^{-k} M^2\right)\right]^{-1} \, \left[-e^{\omega_0^+} (M (\omega_0^+-1)+2 \omega_0^+) \left((M+1) \omega_0^- \right. \right. \nonumber \\
   \left. \left. \left(\frac{1}{6}
   \tilde{\textrm{M}}_{0,\frac{3}{2}}\left(-2 \omega_0^-\right) +R^-\right)    +(\omega_0^--1)
   \tilde{\textrm{M}}_{0,\frac{1}{2}}\left(-2 \omega_0^-\right)\right)  -\left(M R^-+2 e^{-\omega_0^-} \omega_0^-\right)
   \left((M+1) \right. \right. \nonumber \\
   \left. \left. \left(\omega_0^+ \tilde{\textrm{W}}_{0,\frac{3}{2}}\left(-2 \omega_0^+\right)+e^{\omega_0^+} (\omega_0^+-1)^2\right)  +(\omega_0^++1) \tilde{\textrm{W}}_{0,\frac{1}{2}}\left(-2 \omega_0^+\right)\right)  -M e^{\omega_0^-}
   (\omega_0^--1) \right.  \nonumber \\
   \left.  \left((\omega_0^++1) \tilde{\textrm{M}}_{0,\frac{1}{2}}\left(-2 \omega_0^+\right)-(M+1) \omega_0^+
    \left(\frac{1}{6} \tilde{\textrm{M}}_{0,\frac{3}{2}}\left(-2 \omega_0^+\right)+R^+\right)\right)+ \left(M R^++2 e^{\omega_0^+} \omega_0^+\right)  \right.   \nonumber \\
   \left.  \left((M+1) \left(\omega_0^- \tilde{\textrm{W}}_{0,\frac{3}{2}}\left(-2 \omega_0^-\right)+e^{\omega_0^-} (\omega_0^--1)^2\right)   +(1-\omega_0^-) \tilde{\textrm{W}}_{0,\frac{1}{2}}\left(-2 \omega_0^-\right)\right)\right]~,
\end{align}
where $\tilde{\textrm{M}}_{l,n}(x) = \frac{\partial \textrm{M}_{l,n}(x)}{\partial l}$, $\tilde{\textrm{W}}_{l,n}(x) = \frac{\partial \textrm{W}_{l,n}(x)}{\partial l}$, $\omega_0^{\pm} = \omega_0 \pm (k/2)$ and $R^{\pm} = (\omega_0^{\pm}+1)\, e^{-\omega_0^{\pm}}+(\omega_0^{\pm}-1)\, e^{\omega_0^{\pm}}$. The analytic asymptotes in figure \ref{fig8}(b) are generated using this expression, where $\sigma - \sigma(St =0) = \Im(\omega_1\, l)+\textit{O}(l^2)$. Similarly, equations (\ref{eqnD7})-(\ref{eqnD8}) also need to be expanded in $l$ to obtain a correct solution of $\textit{O}(St)$ accurate.

\bibliographystyle{jfm}
\bibliography{jfm}







\end{document}